\documentclass[acmsmall]{acmart}

\usepackage{pgfplots}
\pgfplotsset{compat=newest}
\usepackage[linesnumbered,inoutnumbered,ruled]{algorithm2e}
\usepackage{mathtools}
\usepackage{bm}
\usepackage{multirow}
\usepackage{enumitem}
\usepackage{bbm}
\usepackage{anyfontsize}

\setcopyright{none}
\renewcommand\footnotetextcopyrightpermission[1]{}

\newcommand{\argmax}{argmax}
\newcommand{\argmin}{argmin}
\newcommand{\RR}{\mathbb{R}}
\newcommand{\ZZ}{\mathbb{Z}}
\newcommand{\EE}{\mathbb{E}}
\newcommand{\PP}{\mathbb{P}}

\newcommand{\pr}{^{\prime}}

\newcommand\given[1][]{\:#1\vert\:}

\definecolor{mine}{HTML}{C9D290}

\pgfplotsset{
  layers/axis lines on top/.define layer set={
    axis background,
    axis grid,
    axis ticks,
    axis tick labels,
    pre main,
    main,
    axis lines,
    axis descriptions,
    axis foreground,
  }{/pgfplots/layers/standard},
}

\newcommand{\name}{PPGA\xspace}
\newcommand{\github}{\url{https://github.com/uwaterloo-mast/PPGA}\xspace}

\newcommand{\reffig}[1]{Figure \ref{#1}}
\newcommand{\reftab}[1]{Table \ref{#1}}
\newcommand{\refsec}[1]{Section \ref{#1}}

\newcommand{\refapp}[1]{Appendix \ref{#1}}
\newcommand{\refequ}[1]{(\ref{#1})}
\newcommand{\refequs}[2]{(\ref{#1})--(\ref{#2})}

\newcommand{\refalg}[1]{Algorithm \ref{#1}}
\newcommand{\refthm}[1]{Theorem \ref{#1}}
\newcommand{\reflemm}[1]{Lemma \ref{#1}}

\newcommand{\refclm}[1]{Claim \ref{#1}}

\newcommand{\refdef}[1]{Defenision \ref{#1}}
\newcommand{\refline}[2]{Line \ref{#1} of Algorithm \ref{#2}}

\newtheorem{theorem}{Theorem}
\newtheorem{definition}[theorem]{Definition}
\newtheorem{lemma}[theorem]{Lemma}
\newtheorem{claim}[theorem]{Claim}

\begin{document}

\title{Asymptotically Fair and Truthful Allocation of Public Goods}

\author{Pouya Kananian}
\authornote{This research was conducted while the author was a graduate student at the University of Waterloo.}
\email{pouya.kananian@mail.utoronto.ca}
\affiliation{
       \institution{University of Toronto}
       \city{Toronto}
       \country{Canada}
}

\author{Arnesh Sujanani}
\email{arnesh.sujanani@uwaterloo.ca}
\affiliation{
       \institution{University of Waterloo}
       \city{Waterloo}
       \country{Canada}
}

\author{Seyed Majid Zahedi}
\authornote{Corresponding Author.}
\email{smzahedi@uwaterloo.ca}
\affiliation{
       \institution{University of Waterloo}
       \city{Waterloo}
       \country{Canada}
}

\begin{abstract}
We study the fair and truthful allocation of $m$ divisible public items among $n$ agents, each with distinct preferences for the items.
To aggregate agents' preferences fairly, we focus on finding a core solution.
For divisible items, a core solution always exists and can be calculated by maximizing the Nash welfare objective.
However, such a solution is easily manipulated; agents might have incentives to misreport their preferences.
To mitigate this, the current state-of-the-art finds an approximate core solution with high probability while ensuring approximate truthfulness.
However, this approach has two main limitations.
First, due to several approximations, the approximation error in the core could grow with $n$, resulting in a non-asymptotic core solution.
This limitation is particularly significant as public-good allocation mechanisms are frequently applied in scenarios involving a large number of agents, such as the allocation of public tax funds for municipal projects.
Second, implementing the current approach for practical applications proves to be a highly nontrivial task.
To address these limitations, we introduce \name, a (differentially) Private Public-Good Allocation algorithm, and show that it attains asymptotic truthfulness and finds an asymptotic core solution with high probability.
Additionally, to demonstrate the practical applicability of our algorithm, we implement \name and empirically study its properties using municipal participatory budgeting data.
\end{abstract}

\pagestyle{fancy}
\fancyhead{}
\fancyfoot{}

\maketitle
\thispagestyle{empty}

\section{Introduction}
\label{sec:intro}

Unlike the allocation of private goods, where each item goes to a single agent, public goods allow multiple agents to benefit from an allocated item.
In this paper, we study the problem of fairly allocating $m$ divisible public goods among $n$ agents in a truthful manner.
Different agents hold distinct preferences for the items.
Each item has a size, and the total size of allocated items should not exceed the available capacity.
The fair allocation of divisible public goods is a fundamental problem in social choice theory with many real-world applications.
Examples include:
(1) federal/state budget allocations between services such as healthcare, education, and defense or municipal budget allocations to improve utilities such as libraries, parks, and roads%
\footnote{Municipal budgets are more amenable to inclusive voting procedures like participetory budgeting, while federal/state budgets are more likely voted on by legislators};
(2) shared memory allocations between files with different sizes; and
(3) time allocations between activities during events.

An allocation mechanism produces outcomes based on reported preferences of all agents.
Agents need not reveal their true preferences but strategically report them to maximize their utility.
For instance, consider a setting where there are one or more commonly preferred items.
Such items are highly likely to be allocated regardless of the reported preferences of a single agent.
Given this and assuming that other agents report their preference truthfully, agents could be incentivized to \emph{free-ride} by falsely claiming disinterest in commonly preferred items and reporting preferences only for their individually preferred items.
By doing so, free riders increase the chances of their individually preferred items being allocated under a fair allocation mechanism.

To aggregate agents' preferences fairly, we focus on the classic game theoretic notion of the core~\cite{gillies1953some,debreu1963limit}.
The core generalizes well-studied notions of proportionality and Pareto efficiency by ensuring group-wise fairness, providing fair outcomes to each agent subset relative to its size.
The notion of the core has been extensively studied in the context of public-good allocation~\cite{foley1970lindahl,muench1972core,fain2016core,fain2018fair,munagala2022approximate}.
For allocating divisible public goods, the core always exists, and it can be calculated by maximizing \emph{Nash welfare (NW)} objective (i.e., the product of agents' utilities)~\cite{fain2016core}.
However, the core is easy to manipulate; agents might be incentivized to free-ride.

To address this issue, Fain et al.~\cite{fain2016core} propose a method that aims to find an approximate core solution with high probability while also achieving approximate truthfulness.
This method relies on the exponential mechanism derived from differential privacy~\cite{mcsherry2007mechanism}.
The exponential mechanism uses a scoring function to assign a score to each outcome.
Subsequently, a sample is drawn from a distribution that exponentially weights outcomes based on their scores.
This guarantees that the selected outcome's score is approximately maximized with high probability.

Informally, differential privacy ensures that the output of a mechanism does not change significantly when any agent unilaterally modifies their data.
This emphasis on unilateral deviations aligns closely with truthfulness in mechanism design, where a mechanism is truthful if agents have no incentive to misreport their types.
As a result, differentially private mechanisms can be shown to be approximately dominant-strategy truthful~\cite{mcsherry2007mechanism}.
For the exponential mechanism, the level of differential privacy--and consequently, truthfulness--is contingent on the sensitivity%
\footnote{Informally, the sensitivity of a function is the maximum change in its output resulting from a change in its input (refer to \refsec{subsec:md-vi-dp} for a formal definition).} of the scoring function to the reported input of any individual agent.
Higher sensitivity corresponds to a lower quality of the guarantee.

The use of the exponential mechanism for public-good allocation faces two primary challenges.
First, while the NW objective seems to be an ideal choice for the scoring function, its direct use is hindered by its high sensitivity to each agent's reported preferences.
This limitation arises since the NW objective is not separable%
\footnote{$f(x)$ is separable with respect to a partition of $x$ into $n$ sub-vectors $x = (x_1, \dots, x_n)$ if $f(x) = \sum f_i(x_i)$.}.
To address this, Fain et al.~\cite{fain2016core} propose using a proxy function to replace the NW objective in the scoring function.

The introduced proxy function strikes a balance between reducing the sensitivity of the scoring function--thereby improving truthfulness approximation--and retaining sufficient sensitivity to ensure an acceptable approximation to the core.
However, the use of the proxy function, along with other approximations, introduces an approximation error in satisfying the core conditions.
This approximation error can grow with the number of agents, potentially resulting in a solution that does not satisfy the asymptotic core.
This limitation is particularly significant as public-good allocation mechanisms are frequently applied in scenarios involving a large number of agents, such as participatory budgeting elections for distributing municipal budgets.

Secondly, sampling an $m$-dimensional allocation from a distribution poses a significant practical challenge.
To tackle this, Fain et al.~\cite{fain2016core} propose employing the hit-and-run method~\cite{smith1984efficient} to sample an allocation from an ``approximately right distribution.''
However, implementing the hit-and-run method for practical applications proves to be a highly nontrivial task, as discussed in the conclusion of Sec.\ 2.2 by~\citeauthor{lovasz2007geometry}~\cite{lovasz2007geometry}.
Moreover, the implications of the extra approximation on the guarantees of truthfulness and core remain unclear.

\subsection{Our Contributions}\label{subsec:our-contributions}
In \refsec{sec:algorithm}, we introduce \name, a novel differentially private algorithm for public-good allocation.
A key feature of \name is its approach to maximize the NW objective in a differentially private way without requiring a proxy objective.
As previously discussed, the non-separable nature of the NW objective poses challenges in deploying differentially private mechanisms~\cite{fain2016core}.
To tackle this challenge, we employ a key technique called \emph{global variable consensus optimization}~\cite{boyd2011distributed}.
Consensus transforms the NW objective into a separable form that splits easily.
Leveraging the \emph{alternating direction method of multipliers (ADMM)}~\cite{glowinski1975approximation,gabay1976dual} enables us to maximize the NW objective in a distributed manner.
And this further allows us to employ the \emph{Gaussian mechanism}~\cite{mcsherry2007mechanism} from differential privacy to achieve truthfulness.
The application of differentially private ADMM to the global variable consensus problem for Nash welfare optimization is a novel contribution of this work.

In \refsec{sec:analysis}, we analytically study the properties of our proposed algorithm.
Our primary technical contribution lies in showing that, for carefully chosen $\epsilon, \delta > 0$, \name achieves $(\epsilon, \delta)$-truthfulness and returns an $(\varepsilon_1, \varepsilon_2(1 + \varepsilon_1))$-core outcome with probability at least $1 - 1/n - 1/n^m$, where
\[
    \varepsilon_1 = \mathcal{O}\left(\sqrt{\frac{m\log(1/\delta)}{n\epsilon^2}}\right) \quad \text{and} \quad \varepsilon_2 = \mathcal{O}\left(\sqrt[4]{\frac{m\log(1/\delta)}{n\epsilon^2}}\right).
\]
Assuming that $m = o(\sqrt{n})$ and setting $\epsilon = \Theta(1/\log(n))$ and $\delta = \Theta(1/\sqrt{n})$, we further demonstrate that \name is asymptotically truthful (\refthm{thm:SP}) and yields an asymptotic core solution with high probability (\refthm{thm:asympt-core}).
To our knowledge, \name is the first polynomial-time algorithm (\refthm{thm:comp-comp}) that offers these guarantees.

In \refsec{sec:experiments}, we demonstrate that \name can be deployed in practice to solve large-scale public-good allocation problems.
To this end, we implement \name and utilize our implementation to compare the outcome of \name with a core solution using data obtained from real-world participatory budgeting elections~\cite{faliszewski2023participatory}.
\section{Preliminaries}
\label{sec:problem}
In this section, we first define the public-good allocation problem and its desired properties.
We then provide an overview of differential privacy as a tool for designing truthful mechanisms.
A summary of our notations is presented in~\refapp{sec:notations}

\subsection{Problem Formulation}\label{subsec:formulation}
We consider a public-good allocation problem with $n$ agents and $m$ divisible public items ($m \ll n$).
The size of each item $j$ is denoted by $s_j \in \RR_{>0}$, and the size vector is denoted by $s = (s_1, \dots, s_m)$.
The total available capacity is $c \in \RR_{>0}$.
An allocation is a vector $z = (z_1, \dots, z_m) \in [0,1]^m$, where $z_j$ represents the fraction of the total capacity that is allocated to item $j$.
The set of all feasible allocations is denoted by:
\[
  \mathcal{Z} \triangleq \{z \in [0, 1]^m \given \Vert z \Vert_1 \le 1, \; cz \le s \}.
\]
Agent $i$'s utility function for an allocation $z \in \mathcal{Z}$ is denoted by $U_i(z)$ and is parameterized by the utility vector $u_i = (u_{i1}, \dots, u_{id})$, where $d$ is a positive integer.
For example, for a linear utility function of the form $U_i(z) = \sum_{j=1}^m u_{ij} z_j$, we have $d = m$, and each $u_{ij}$ represents the relative value that agent $i$ assigns to the fraction of the budget allocated to item $j$.
In this paper, we focus on a subclass of utility functions defined on $\mathcal{Z}$ that are differentiable, strictly increasing, concave, and $\beta$-smooth, i.e., they have $\beta$-Lipschitz continuous gradients:
\[
 \Vert \nabla U_i(z) - \nabla U_i(z\pr) \Vert_2 \le \beta \Vert z - z\pr \Vert_2.
\]
This subclass includes the common linear utility functions, which generalize additive utilities studied by~\cite{bogomolnaia2005collective,fain2018fair,bhaskar2018truthful,aziz2018proportionally,talmon2019framework,michorzewski2020price,brandl2021distribution,peters2021proportional,benade2021preference}.
Without loss of generality, we assume that $U_i \in [0,1]$ for all $i$, with $U_i(\bm{0}_m) = 0$ and $U_i(z) > 0$ for some $z \in \mathcal{Z}$.
We further assume that $u_i \in \mathcal{U}$ for every $i$, where $\mathcal{U} \triangleq [0,1]^d$.

\subsection{Mechanism Design for Public Goods}\label{subsec:md-for-pg}
A randomized allocation mechanism $M$ produces a probability distribution over feasible allocations given agents' reported utilities $u = (u_1, \dots, u_n) \in \mathcal{U}^n$.
We use $M(u)$ to denote the distribution produced by mechanism $M$ for the reported utilities $u$, and at times, we also use $M(u)$ to represent a random allocation drawn from the distribution $M(u)$, slightly abusing the notation.
Agents need not report their true utilities.
They report strategically to optimize their total utility taking into account what (they think) other agents report.
If agents are always incentivized to report their true utilities, no matter what others do, then the mechanism is \emph{dominant-strategy truthful}:
\begin{definition}[Dominant-strategy Truthfulness]
 Let $U_i$ be agent $i$'s utility function parameterized by $i$'s true utility vector $u_i$.
 A randomized mechanism $M$ is $(\epsilon,\delta)$-truthful if $\EE[U_i(M(u_i, u_{-i}))] \ge (1-\epsilon)\EE[U_i(M(u\pr_i, u_{-i}))] - \delta$ for every $i$, $u\pr_i \in \mathcal{U}$, and $u_{-i} \in \mathcal{U}^{(n-1)}$.%
 \footnote{Subscript $-i$ is used to refer to all agents other than agent i.}
\end{definition}

If $\epsilon, \delta = 0$, then $M$ is \emph{exactly truthful}.
Approximate truthfulness is desirable in settings in which the approximation parameters $\epsilon$ and $\delta$ tend to 0 as the number of agents $n$ grows large.
This property is referred to as \emph{asymptotic truthfulness}.
Next, we formally define the classic notion of the \emph{core}.

\begin{definition}[Core]
\label{def:core}
 For an allocation $z \in \mathcal{Z}$, a set of agents $A$ form a blocking coalition if there exists another allocation $z\pr \in \mathcal{Z}$ such that $(\vert A \vert/n) U_i(z\pr) \geq U_i(z)$ for every $i \in A$ with at least one strict inequality.
 An allocation is a core outcome if it admits no blocking coalitions.
\end{definition}
\noindent
In this definition, when a subset $A$ of agents deviates, they can choose any feasible allocation with the full capacity $c$.
However, their utility is scaled down by a factor of $\vert A \vert/n$.
An alternative way of defining a core solution is where a deviating coalition $A$ could choose any allocation with a capacity of $(c\vert A \vert)/n$ instead of $c$, but their utilities would not be scaled down~\cite{foley1970lindahl,scarf1967core}.
For $\vert A \vert = n$, both notions capture Pareto efficiency.
However, for $\vert A \vert = 1$, they provide different interpretations of proportionality--one based on utility and one based on capacity.

For divisible goods, the core coincides with the \emph{max Nash welfare (MNW)} solution:%
\footnote{Similar lemmas appear in~\cite{fain2016core,fain2018fair} for other classes of utility functions.
For completeness, we provide the proof of \reflemm{lemm:core-solution} in \refapp{subsec:proof-lemm-core-solution}.}
\begin{lemma}
\label{lemm:core-solution}
 If each $U_i$ is differentiable and concave, then any allocation that maximizes $\sum_i \log(U_i(z))$ subject to $z \in \mathcal{Z}$ constitutes a core solution%
 \footnote{In this paper, all logarithms are natural.}.
\end{lemma}
\noindent
This lemma shows that the exact MNW solution is a core outcome.
However, such a solution can be irrational even when all inputs are rational~\cite{airiau2018positional}, potentially precluding the existence of an exact algorithm~\cite{fain2018fair}.
Therefore, we adopt an approximate notion of the core that still provides meaningful guarantees:

\begin{definition}[Approximate core]
 For $\epsilon, \delta \ge 0$, an allocation $z \in \mathcal{Z}$ is an $(\epsilon, \delta)$-core outcome if there exists no set of agents $A \subseteq N$ and no allocation $z\pr \in \mathcal{Z}$ such that $(\vert A \vert/n)U_i(z\pr) \ge (1 + \epsilon) U_i(z) + \delta$ for all $i \in A$ with at least one strict inequality.
\end{definition}
\noindent
When $\epsilon$ and $\delta$ converge to zero asymptotically as $n$ grows large, the allocation is said to be an \emph{asymptotic core} solution.
The following lemma shows that an approximate MNW solution implies an approximate core solution (see \refapp{subsec:proof-lemm-mnw-core-approx} for the proof).

\begin{lemma}
\label{lemm:mnw-core-approx}
 Let $\epsilon, \delta \ge 0$.
 Then, $z \in \mathcal{Z}$ is an $(\epsilon, \delta)$-core outcome if, for any $z\pr \in \mathcal{Z}$, we have:
 \begin{equation}
 \label{eq:sum-tilde-z}
  \frac{1}{n}\sum_i \frac{U_i(z\pr)}{U_i(z) + \delta/(1 + \epsilon)} \le 1 + \epsilon
 \end{equation}
\end{lemma}

\subsection{Mechanism Design via Differential Privacy}\label{subsec:md-vi-dp}
In this subsection, we provide some background on differential privacy as a tool for designing truthful mechanisms.
Informally, a mechanism satisfies DP if its output is nearly equally likely to be observed for any pair of \emph{adjacent} inputs.
Inputs are considered adjacent if they differ in only one element.
For allocation mechanisms, inputs correspond to agents' reported utilities.
Thus, $u, u\pr \in \mathcal{U}^n$ are adjacent if they differ solely in the reported utility of a single agent.
We now formally define DP~\cite{DP06}:
\begin{definition}[DP]
 A randomized mechanism $M$ is $(\epsilon, \delta)$-DP if, for any two adjacent inputs $u, u\pr \in \mathcal{U}^n$ and any subset of outputs $O \subseteq \mathcal{Z}$, it satisfies $\PP[M(u) \in O] \le e^{\epsilon} \PP[M(u\pr) \in O] + \delta$.%
 \footnote{Symmetry of adjacency relation implies: $\PP[M(u) \in O] \ge e^{-\epsilon}\PP[M(u\pr) \in O] - e^{-\epsilon} \delta \ge e^{-\epsilon}\PP[M(u\pr) \in O] - \delta$.}
\end{definition}
\noindent
In this definition, $\epsilon$ and $\delta$ control the desired level of privacy and are typically provided as inputs to the mechanism.
In general, smaller values provide stronger privacy guarantees but result in higher levels of noise being required to be injected, which can adversely affect the quality of the output.
A mechanism that satisfies $(\epsilon, \delta)$-DP is $(\epsilon, \delta)$-truthful:%
\footnote{When $\delta = 0$, McSherry and Talwar~\cite{mcsherry2007mechanism} show that mechanisms satisfying $\epsilon$-differential privacy make truth-telling an $(\exp(\epsilon) - 1)$-approximately dominant strategy.
However, we are not aware of any existing result for the case $\delta > 0$.
Therefore, for completeness, we provide a proof of \reflemm{lemm:dp-2-sp}.}
\begin{lemma}
\label{lemm:dp-2-sp}
 Let $M$ be $(\epsilon, \delta)$-DP for some $\epsilon, \delta < 1$.
 Then, $M$ is $(\epsilon, \delta)$-truthful.
\end{lemma}

\begin{proof}
 Consider any agent $i$, and let $U_i: \mathcal{Z} \mapsto [0,1]$ be agent $i$'s utility parameterized according to their true utility vector $u_i$.
 Define the set $S(t) = \{ z \given U_i(z) > t \}$.
 Since $M$ is $(\epsilon, \delta)$-DP, for any $u = (u_i, u_{-i}) \in \mathcal{U}^n$ and $u_i\pr \in \mathcal{U}$, the following inequality holds:
 \begin{equation}
 \label{eq:dp2sp1}
  \PP[M(u) \in S(t)] \ge e^{-\epsilon} \PP[M(u^\prime_i, u_{-i}) \in S(t)] - \delta.
 \end{equation}
 Given the definition of $S(t)$, we can rewrite \refequ{eq:dp2sp1} as:
 \begin{equation}
 \label{eq:dp2sp2}
  \PP[U_i(M(u)) > t] \ge e^{-\epsilon} \PP[U_i(M(u^\prime_i, u_{-i})) > t] - \delta.
 \end{equation}
 Given that $\EE[X]=\int_0^1 \PP[X> t] dt$ for any random variable $X \in [0, 1]$, we obtain the following by integrating both sides of \refequ{eq:dp2sp2}:
 \begin{align*}
  \EE[U_i(M(u))] \ge e^{-\epsilon}\EE[U_i(M(u\pr_i,u_{-i}))] - \delta \ge (1-\epsilon) \EE[U_i(M(u\pr_i,u_{-i}))] - \delta,
 \end{align*}
 where the second inequality follows because $e^{-\epsilon} \ge 1 - \epsilon$.
\end{proof}

\noindent
We next define \emph{R{é}nyi differential privacy (RDP)} as a relaxation of DP~\cite{RDP17}:
\begin{definition}[RDP]
 A randomized mechanism $M$ is $(\alpha, \epsilon)$-RDP with order $\alpha > 1$ if for any two adjacent inputs $u,u\pr \in \mathcal{U}^n$, it satisfies: $D_\alpha(M(u) \Vert M(u\pr)) \le \epsilon$, where $D_\alpha$ is the R{é}nyi divergence of order $\alpha$ defined as:
 \[
  D_\alpha(P \Vert Q) \triangleq \frac{1}{\alpha - 1} \log\left(\EE_{X\sim Q}\left[\left(\frac{P(X)}{Q(X)}\right)^\alpha\right]\right).
 \]
\end{definition}

RDP provides strong guarantees regarding the concept of sequential \emph{composition}.
If $M_1$ and $M_2$ are $(\alpha,\epsilon_1)$-RDP and $(\alpha, \epsilon_2)$-RDP, respectively, then the mechanism $M_{1,2}$ defined as $M_{1,2}(x) \triangleq (M_1(x), M_2(x))$ is $(\alpha,\epsilon_1 + \epsilon_2)$-RDP~\cite[Proposition 1]{RDP17}.
This property enables straightforward tracking of cumulative privacy loss for iterative mechanisms.
If each iteration of an iterative mechanism is $(\alpha,\epsilon)$-RDP, then $K$ iterations of the mechanism are $(\alpha,K\epsilon)$-RDP\@.
We use this property to analyze our proposed mechanism in \refsec{subsec:asymptotic-sp}.

A common tool for achieving RDP is the \emph{Gaussian mechanism}.
The Gaussian mechanism evaluates a vector-valued function on the input and adds Gaussian noise independently to each coordinate of the output.
The noise magnitude is calibrated to the function's $\ell_2$ sensitivity.
\begin{definition}[L2 sensitivity]
 The $\ell_2$ sensitivity of $f: \mathcal{U}^n \mapsto \RR^m$ is defined as:
 \[
  \Delta_2(f) \triangleq \max_{\text{adj } u,u\pr \in \mathcal{U}^n}\Vert f(u) - f(u\pr) \Vert_2.
 \]
\end{definition}
\noindent
Given this definition, the Gaussian mechanism is formally defined as follows.
\begin{definition}[Gaussian mechanism]
 Let $\mathcal{N}(\mu,\Sigma)$ denote a multivariate normal distribution with mean vector $\mu$ and covariance matrix $\Sigma$.
 For $\alpha > 1$, $\epsilon > 0$, and function $f: \mathcal{U}^n \mapsto \RR^m$ with an $\ell_2$ sensitivity of $\Delta_2(f)$, the Gaussian mechanism $M^G_{f, \alpha, \epsilon}$ is defined as:
 \[
  M^G_{f, \alpha, \epsilon}(u) \triangleq \mathcal{N}(f(u),\sigma^2 I_m),
 \]
 where $I_m$ is the $m\times m$ identity matrix, and $\sigma^2 = \alpha \Delta_2^2(f)/2\epsilon$.
\end{definition}
\noindent
$M^G_{g,\alpha, \epsilon}$ is $(\alpha, \epsilon)$-RDP~\cite[Corollary 3]{RDP17}.
Moreover, if a mechanism is $(\alpha, \epsilon)$-RDP, then it is $(\epsilon+\log(1/\delta)/(\alpha - 1), \delta)$-DP for any $0 < \delta < 1$~\cite[Proposition 3]{RDP17}.
\section{Algorithm}
\label{sec:algorithm}

In this section, we present \name{} (\refalg{alg:PPGA}), an algorithm that directly maximizes a \emph{smoothed} version of the NW objective in a DP manner.
Our approach involves a transformation of the objective into a \emph{separable} form.
Initially, we reframe the optimization problem of \reflemm{lemm:core-solution} into a consensus problem.
Next, we convert the consensus problem into a distributed optimization using ADMM\@.
Finally, to ensure truthfulness, we deploy the Gaussian mechanism.

\subsection{Distributed Maximization of Nash Welfare}\label{subsec:dist-mnw}

The NW objective function, $\sum_i \log(U_i(z))$, poses two challenges.
First, it is undefined when any agent receives zero utility.
Second, it is non-separable, as the shared variable $z$ appears in all terms.
To address the first issue, we use a smooth version of the NW objective: $\sum_i \log(U_i(z) + \upsilon)$, where $\upsilon > 0$ is a small smoothing parameter that vanishes asymptotically as the number of agents increases.
To tackle the second issue, we introduce local variables $x_i$ for each agent and a shared global variable $z$:
\begin{equation}
\label{eq:modified-opt}
 \begin{aligned}[t]
  &\text{Max.} &&\theta(x), && \\
  &\textrm{s.t.} &&z = x_i && \forall i \in 1, \dots, n, \\
  & &&x_i \in \mathcal{Z} && \forall i \in 1, \dots, n,
 \end{aligned}
\end{equation}
\noindent
where $\theta(x)$ is defined for $x = (x_1,\dots,x_n)$ as:
\[
 \theta(x) \triangleq \sum_i \theta_i(x_i) = \sum_i \log(U_i(x_i) + \upsilon).
\]
This is referred to as the \emph{global variable consensus problem}, as it requires all local variables to reach agreement by being equal.
Consensus transforms the additive objective, which does not split, into a separable objective, which splits easily.

The partial augmented Lagrangian~\cite{hestenes1969multiplier,powell1969method} for \refequ{eq:modified-opt} is defined as:
\begin{align*}
 L^\rho(x, z, \gamma) \triangleq \sum_i L_i^\rho(x_i, z, \gamma_i) = \sum_i \left( \theta_i(x_i) - \gamma_{i}^T (x_{i} - z)   - \frac{\rho}{2} \Vert x_{i} - z \Vert_2^2 \right),
\end{align*}
\noindent
where $\gamma_i$ is a dual variable corresponding to the constraint $z = x_i$, and $\rho > 0$ is a penalty parameter.
Note that, similar to $\theta$, the function $L^\rho$ is separable in $x$ and splits into separate components $L_i^\rho$ for each agent $i$.
We next apply ADMM to solve \refequ{eq:modified-opt} in a distributed way through the following iterative updates:
\begin{subequations}
\label{eq:def-standard-admm}
 \begin{align}
  &x_i^{(k)} \coloneqq \underset{x_i \in \mathcal{Z}}{\argmax} \; L_i^\rho(x_i, z^{(k-1)}, \gamma^{(k-1)}_i) &\forall i \in 1, \dots, n, \label{eq:def-standard-admm1}\\
  &z^{(k)} \coloneqq \underset{z}{\argmax} \; L^\rho(x^{(k)}, z, \gamma^{(k-1)}), \label{eq:def-standard-admm2}\\
  &\gamma_{i}^{(k)} \coloneqq \gamma_{i}^{(k-1)} + \rho (x_{i}^{(k)} - z^{(k)}) &\forall i \in 1, \dots, n. \label{eq:def-standard-admm3}
 \end{align}
\end{subequations}

In \refequ{eq:def-standard-admm1}, $x_i^{(k)}$'s can be computed independently for each agent $i$.
Moreover, we can solve \refequ{eq:def-standard-admm2} exactly by setting the gradient $\partial L^\rho / \partial z = \sum_i \left(\gamma^{(k-1)}_{i} + \rho (x^{(k)}_{i} - z^{(k)})\right)$ to zero, which leads to the following closed-form solution:
\begin{equation}
\label{eq:def-modified-admm2}
 z^{(k)} = \frac{1}{n}\sum_i x_{i}^{(k)} + \frac{1}{n\rho}\sum_i \ \gamma_{i}^{(k-1)}.
\end{equation}
We can find an optimal solution to \refequ{eq:modified-opt} through ADMM's iterative updates.
However, this procedure is not truthful.
To address this limitation, we next incorporate DP into the process as a means of achieving truthfulness.

\subsection{DP for Maximizing Nash Welfare}\label{subsec:dp-for-mnw}
To illustrate our proposed mechanism, it might be beneficial to interpret ADMM as an interactive process.
At iteration $k$, each agent $i$ calculates the local variable $x_i^{(k)}$ autonomously.
Given $z^{(k-1)}$ and $\gamma_i^{(k-1)}$, the value of $x_i^{(k)}$ depends solely on agent $i$'s own utility.
With $x_i^{(k)}$ and $z^{(k)}$ known, each agent $i$ independently calculates $\gamma_i^{(k)}$.
These local variables are then submitted by agents, aggregated by the algorithm, and used to compute the global variable $z^{(k)}$.
This resultant global variable is broadcast back to the agents for the next iteration.

In the context of this interactive process, to ensure DP, it is imperative that the value of the global variable remains insensitive to any individual local variable.
To achieve this, we employ the Gaussian mechanism, adding a normal random vector $q^{(k)}$ to $z^{(k)}$:
\begin{equation}
\label{eq:def-noisy-admm2}
 z^{(k)} = \frac{1}{n}\sum_i x_{i}^{(k)} + \frac{1}{n\rho}\sum_i \ \gamma_{i}^{(k-1)} + q^{(k)}.
\end{equation}
According to \refequ{eq:def-standard-admm3}, we have:
\begin{equation}
\label{eq:def-noisy-admm2-2nd-term}
 \sum_i \gamma_{i}^{(k)} = \sum_i \left( \gamma_{i}^{(k-1)} + \rho (x_{i}^{(k)} - z^{(k)})\right).
\end{equation}
Replacing $z^{(k)}$ from \refequ{eq:def-noisy-admm2} into \refequ{eq:def-noisy-admm2-2nd-term}, we get $\sum_i \gamma_{i}^{(k)} = - \rho n q^{(k)}$, which is used to rewrite \refequ{eq:def-noisy-admm2} as:
\begin{equation}
\label{eq:def-new-noisy-admm2}
 z^{(k)} = \frac{1}{n}\sum_i x_{i}^{(k)} - q^{(k-1)} + q^{(k)}.
\end{equation}
This update rule shows how $z^{(k)}$ can be calculated by adding Gaussian noise to the average of $x_i^{(k)}$'s.
The magnitude of the noise can be adjusted to achieve a desired DP guarantee.

\begin{algorithm}[!t]
\SetKwInput{Output}{Output}%
\SetKwInput{Params}{Parameters}%
 \Params{$K \in \ZZ$, $\upsilon, \epsilon, \delta \in (0,1)$, $\alpha > 1$}
 $\epsilon\pr \gets (1/K)(\epsilon - \log(1/\delta)/(\alpha - 1))$\;
 $\sigma^2 \gets \alpha/n^2\epsilon^\prime$\;
 $q^{(0)}, z^{(0)},\gamma^{(0)}_i, x^{(0)}_i = \bm{0_m}$ $\quad \forall i \in 1, \dots, n$\;
 \For{$k = 1, \dots, K$}{
  $x_i^{(k)} \gets \argmax_{x_i \in \mathcal{Z}}(L_i^\rho(x_i, z^{(k-1)}, \gamma^{(k-1)}_i))$ $\quad \forall i \in 1, \dots, n$\;\label{alg:x-update}
  $q^{(k)} \sim \mathcal{N}(0,\sigma^2 I_m)$\;\label{alg:v-update}
  $z^{(k)} \gets (1/n)\sum_i x_{i}^{(k)} + q^{(k)} - q^{(k-1)}$\;\label{alg:z-update}
  $\gamma_{i}^{(k)} \gets \gamma_{i}^{(k-1)} + \rho (x_{i}^{(k)} - z^{(k)})$ $\quad \forall i\in 1, \dots, n$\;\label{alg:g-update}
 }
 $\bar{z} \gets (1/K)\sum_{k = 1}^{K}z^{(k)}$\;
 $\hat{z} \gets \Pi_{\mathcal{Z}}(\bar{z})$\;\label{alg:hat-z}
 \Output{$\hat{z}$}
\caption{Private public-good allocation (\name)}
\label{alg:PPGA}
\end{algorithm}

\refalg{alg:PPGA} shows the pseudocode of our proposed (differentially) private public-good allocation mechanism, \name.
The algorithm takes as parameters $K$, $\upsilon$, $\epsilon$, $\delta$, and $\alpha$.
$K$ specifies the number of iterations.
$\upsilon$ controls the smoothness of the objective function.
$\epsilon$, $\delta$, and $\alpha$ together determine the desired level of privacy--and, consequently, the level of truthfulness.
Specifically, $\epsilon$ and $\delta$ define the level of DP, while $\alpha$ controls the variance of the Gaussian noise (see \refthm{thm:SP}).

At each iteration $k$, the optimal allocation $x_i^{(k)}$ is computed for each agent $i$, given $\gamma_i^{(k-1)}$ and $z^{(k-1)}$.
This step can be executed in parallel for all agents.
The algorithm then computes $z^{(k)}$ as a noisy average of the $x_i^{(k)}$'s.
Given $z^{(k)}$ and $x_i^{(k)}$, the value $\gamma_i^{(k)}$ is then computed for each agent for the next iteration.
After $K$ iterations, the algorithm calculates $\bar{z}$, the time average of the $z^{(k)}$'s, and returns $\hat{z}$, the Euclidean projection of $\bar{z}$ onto $\mathcal{Z}$.%
\footnote{$\Pi_{\mathcal{Z}}(z) = \argmin_{z\pr\in \mathcal{Z}}\Vert z - z\pr \Vert_2^2$.}

\subsection{Discussion}\label{subsec:discussion}

The integration of DP into ADMM inherently presents a trade-off between accuracy and privacy (truthfulness).
Achieving a more accurate MNW solution requires a higher number of iterations.
Fixing the amount of privacy loss per iteration, a higher number of iterations means a higher cumulative privacy loss, resulting in a weaker privacy guarantee.
On the other hand, achieving a stronger privacy guarantee requires a lower cumulative privacy loss.
Fixing the number of iterations, a lower cumulative privacy loss means a higher level of noise per iteration, resulting in diminished accuracy.

The expected value of the noise magnitude at each iteration of \refalg{alg:PPGA} is:
\[
 \EE\left[\Vert q^{(k)} \Vert^2_2\right] = m\sigma^2 = \frac{K m \alpha}{n^2(\epsilon - \log(1/\delta)/(\alpha - 1))}.
\]
Assuming that $m = o(\sqrt{n})$, if we choose $K = \Theta(n)$, $\epsilon = \Theta(1/\log(n))$, and $\delta = \Theta(1/\sqrt{n})$, and set $\alpha = 2\log(1/\delta)/\epsilon + 1$, then the expected noise magnitude at each iteration converges to zero as $n$ grows large--an essential property for achieving an asymptotic core outcome (see \refsec{subsec:asymptotic-core}).

As a final remark, even though we described the algorithm as an interactive process in \refsec{subsec:dp-for-mnw}, we emphasize that our proposed algorithm is neither online nor interactive.
All computations are carried out by the algorithm itself, rather than by the agents.
Agents submit their private utility vectors and, at the end, observe a final allocation vector.
As we show in \refsec{sec:analysis}, the algorithm satisfies DP, ensuring that agents' data remains private.
Moreover, our mechanism guarantees asymptotic truthfulness, meaning that as $n$ grows, agents have no incentive to misreport their utilities.
\section{Analysis}
\label{sec:analysis}
In this section, we first show that \refalg{alg:PPGA} guarantees asymptotic truthfulness.
We then demonstrate that it produces an asymptotic core solution with high probability.
Finally, we analyze its computational complexity.
All omitted proofs are provided in \refapp{sec:app}.

\subsection{Asymptotic Truthfulness}\label{subsec:asymptotic-sp}
To analyze the end-to-end privacy guarantee of \refalg{alg:PPGA}, we separately analyze the DP guarantee of each iteration.
Leveraging the properties of the Gaussian mechanism, we show that each iteration of the algorithm ensures $(\alpha, \epsilon\pr)$-RDP\@.
With the additivity property of RDP~\cite[Proposition 1]{RDP17}, after $K$ iterations, \refalg{alg:PPGA} achieves $(\alpha, K\epsilon\pr)$-RDP\@.
It then follows from~\cite[Proposition 3]{RDP17} that \refalg{alg:PPGA} is $(\epsilon, \delta)$-DP\@.

\begin{lemma}\label{lemm:ppga-dp}
 \refalg{alg:PPGA} is $(\epsilon, \delta)$-DP\@.
\end{lemma}

\begin{proof}
 \refalg{alg:PPGA} consists of $K$ iterations.
 At each iteration $k$, the private data is $x^{(k)}$, while the publicly released data is $z^{(k)}$.
 Note that $\gamma^{(k)}$ is not publicly released, as each $\gamma_i^{(k)}$ is privately computed for each agent $i$.
 The $z$-update step at \refline{alg:z-update}{alg:PPGA} directly applies the Gaussian mechanism to the function $f(x) = \frac{1}{n} \sum_i x_i$.
 Let $x$ and $x^\prime$ be two adjacent inputs that differ only in their $i^{\text{th}}$ element, i.e., $x_i \ne x_i^\prime$.
 Then, we have:
 \[
  \Vert f(x) - f(x\pr)\Vert_2 = \frac{1}{n} \Vert x_i - x_i\pr \Vert_2.
 \]
 Since $x_i, x_i\pr \in [0,1]^m$ and $\Vert x_i \Vert_1, \Vert x_i\pr \Vert_1 \leq 1$, it follows that:
 \begin{equation}
  \label{eq:bounded-domain}
  \Vert x_i - x_i\pr \Vert_2 \leq (\Vert x_i \Vert_2^2 + \Vert x_i\pr \Vert_2^2)^{1/2} \leq \sqrt{2}.
 \end{equation}
 This implies $\Delta_2(f) \leq \sqrt{2}/n$.
 By~\cite[Corollary 3]{RDP17}, each iteration $k$ of the algorithm is $(\alpha, \epsilon\pr)$-RDP\@.
 Consequently, by~\cite[Proposition 1]{RDP17}, the composition of the $K$ iterations satisfies $(\alpha, \bar{\epsilon})$-RDP, where $\bar{\epsilon} = K\epsilon\pr = \epsilon - \log(1/\delta)/(\alpha - 1)$.
 Finally, by~\cite[Proposition 3]{RDP17}, the $K$ iterations of \refalg{alg:PPGA} satisfy $(\epsilon,\delta)$-DP\@.
 It is important to note that computing $\bar{z}$ after the $K$ iterations and projecting it onto $\mathcal{Z}$ are merely post-processing steps.
 Since DP is immune to post-processing~\cite[Proposition 2.1]{DPBook14}, these steps do not affect the privacy guarantees%
 \footnote{If $M$ is $(\epsilon, \delta)$-DP, then applying any randomized mapping $f$ to $M(u)$ preserves the $(\epsilon, \delta)$-DP property.}.
\end{proof}

\noindent
We next establish our first technical result:
\begin{theorem}\label{thm:SP}
 \refalg{alg:PPGA} is asymptotically truthful.
\end{theorem}
\begin{proof}
 By \reflemm{lemm:ppga-dp}, \refalg{alg:PPGA} is $(\epsilon, \delta)$-DP.
 It then follows directly from \reflemm{lemm:dp-2-sp} that it is also $(\epsilon, \delta)$-truthful.
 Setting $\delta = \Theta(1/\sqrt{n})$ and $\epsilon = \Theta(1/\log(n))$, we conclude that \refalg{alg:PPGA} is asymptotically truthful.
\end{proof}

\subsection{Asymptotic Core}\label{subsec:asymptotic-core}
Let $x = (x_1, \dots, x_n)$ and $\gamma = (\gamma_1, \dots, \gamma_n)$, and define $w \triangleq (x, z, \gamma) \in \mathcal{W} \triangleq (\mathcal{Z}^n, \RR^m, \RR^{mn})$.
Let $w^{(k)} \triangleq (x^{(k)}, z^{(k)}, \gamma^{(k)})$, and define $G \triangleq (I_m, \dots, I_m)$.
To show that $\hat{z}$ is an approximate core solution, we aim to derive an upper bound on $\underset{z \in \mathcal{Z}}{\max} ~ \sum_i \frac{U_i(z)}{U_i(\hat{z}) + \upsilon}$.
To this end, we proceed in three steps.
First, we bound $\underset{z \in \mathcal{Z}}{\max} ~ \sum_i \frac{U_i(z)}{U_i(\bar{x}_i) + \upsilon}$, where $\bar{x} = \frac{1}{K}\sum_k x^{(k)}$.
Second, we establish a bound on the distance between $\hat{z}$ and each $\bar{x}$.
Finally, using the smoothness of $U_i$'s and applying \reflemm{lemm:mnw-core-approx}, we conclude that $\hat{z}$ is an approximate core solution.

\begin{lemma}\label{lemm:core-1}
 Let $\{w^{(k)}\}$ and $\{q^{(k)}\}$ be sequences generated by \refalg{alg:PPGA}.
 Then, we have:
 \begin{align}\label{eq:core-1}
  \frac{1}{n}\sum_i\frac{U_i(z)}{U_i(\bar{x}_i) + \upsilon} \le 1 + \frac{\rho}{K}\sum_{k=1}^{K}(z^{(k)} - z)^T q^{(k)} + \frac{\rho}{2K} \quad \quad \forall z \in \mathcal{Z}.
 \end{align}
\end{lemma}

\noindent
The next lemma provides an upper bound on the distance between $\hat{z}$ and any $\bar{x}_i$.

\begin{lemma}\label{lemm:residual}
 Let $\{w^{(k)}\}$ and $\{q^{(k)}\}$ be sequences generated by \refalg{alg:PPGA}.
 Let $w^* = (x^*, z^*, \gamma^*)$ be an optimal solution to \refequ{eq:modified-opt}, with $x^*_i = z^*$ for all $i$.
 Then, we have:
 \begin{align}\label{eq:residual}
  \Vert \bar{x} - G\hat{z} \Vert_2 \le \frac{2\sqrt{n}}{K} \left(\sum_{k=1}^{K} \vert (z^{(k)} - z^*)^T q^{(k)} \vert \right)^{\frac{1}{2}} + \frac{\sqrt{2n}}{K} + \frac{2}{\rho K} \Vert \gamma^* \Vert_2.
 \end{align}
\end{lemma}

The right-hand side of \refequ{eq:core-1} and \refequ{eq:residual} involves random variables--specifically, the sequences $\{z^{(k)}\}$ and $\{q^{(k)}\}$.
The next lemma provides a bound on their tail behavior:

\begin{lemma}\label{lemm:zkqk}
 Let $\{w^{(k)}\}$ and $\{q^{(k)}\}$ be sequences generated by $K = \Theta(n)$ iterations of \refalg{alg:PPGA}.
 Suppose $\epsilon$, $\delta$, and $\alpha$ are chosen such that $m\sigma^2 < 1$.
 Then, for any $z \in \mathcal{Z}$ and some constant $C > 0$, with probability at least $1 - 1/n - 1/n^m$, we have:
 \begin{equation}\label{eq:zkqk}
  \frac{1}{K}\sum_{k = 1}^{K} \left| (z^{(k)} - z)^T q^{(k)} \right| \le C\sqrt{m}\sigma.
 \end{equation}
\end{lemma}

\noindent
Next, we establish that the outcome of \refalg{alg:PPGA} is an approximate core solution:

\begin{lemma}\label{lemm:approx-core}
 Suppose that $m=o(\sqrt{n})$.
 Then, after $K = \Theta(n)$ iterations, \refalg{alg:PPGA} returns an $(\varepsilon_1, \varepsilon_2(1 + \varepsilon_1))$-core outcome with probability at least $1 - 1/n - 1/n^m$, where:
 \[
  \varepsilon_1 = \mathcal{O} \left( \sqrt{\frac{m\log(1/\delta)}{n\epsilon^2}} \right) \quad \text{and} \quad \varepsilon_2 = \mathcal{O}\left( \sqrt[4]{\frac{m\log(1/\delta)}{n\epsilon^2}} \right).
 \]
\end{lemma}

\noindent
Finally, we establish \refalg{alg:PPGA}{}'s asymptotic fairness.

\begin{theorem}\label{thm:asympt-core}
 Suppose that $m = o(\sqrt{n})$.
 Then, after $K = \Theta(n)$ iterations, the output of \refalg{alg:PPGA} is an asymptotic core outcome with probability at least $1 - 1/n - 1/n^m$.
\end{theorem}

\begin{proof}
 If we choose $\upsilon = \Theta(1/\sqrt{n})$, $\delta = \Theta(1/\sqrt{n})$, and $\epsilon = \Theta(1/\log(n))$, and set $\alpha = 2\log(1/\delta)/\epsilon + 1$, then, by \reflemm{lemm:approx-core}, \refalg{alg:PPGA} returns an asymptotic core outcome.
\end{proof}

\subsection{Computational Complexity}\label{subsec:comp-comp}

The main computation in each iteration $k$ of \refalg{alg:PPGA} is to compute $x_i^{(k)}$ for each agent $i$.
This step involves solving a convex program.
Several methods exist for solving broad classes of convex optimization problems with a number of operations that grow polynomially in the problem dimensions and logarithmically in $1/\xi$, where $\xi > 0$ denotes the desired accuracy~\cite{nesterov1994interior}.
Typically, such accuracy guarantees are provided with respect to the objective function value.
For the subproblem in \refline{alg:x-update}{alg:PPGA}, however, we require an accuracy guarantee on the first-order optimality condition (see \refequ{eq:opt-x}).
Fortunately, due to the smoothness of $U_i$, such a guarantee can still be achieved in polynomial time using first-order methods--such as the one proposed by Lu and Mei~\cite{lu2023accelerated}.

\begin{lemma}\label{lemm:x-step-approx}
 For any $x_i \in \mathcal{Z}$ and $z, \gamma_i \in \RR^m$, a point $x_i^* \in \mathcal{Z}$ that satisfies the inequality
 \begin{equation} \label{eq:x-step-approx}
  (x_i - x_i^*)^T ( \nabla \theta_i(x_i^*) - \gamma_i - \rho(x_i^* - z) ) \le \xi
 \end{equation}
 can be computed in time $\mathcal{O}( m \log(m) \sqrt{(L+\rho)/\xi} \log(1/\xi) )$, where $L$ is $\nabla \theta_i$'s Lipschitz constant.
\end{lemma}

\noindent
We now establish our final technical result.

\begin{theorem}\label{thm:comp-comp}
 \refalg{alg:PPGA} achieves asymptotic truthfulness and computes an asymptotic core solution with high probability in polynomial time.
\end{theorem}

\begin{proof}
 Consistent with the proof of \refthm{thm:asympt-core}, let $K = \Theta(n)$ and $\upsilon = \Theta(1/\sqrt{n})$.
 Then, $L \le \frac{(1+\beta)^2}{2\upsilon^2} + \frac{\beta}{\upsilon} = \Theta(n)$.
 By \reflemm{lemm:x-step-approx}, at each iteration $k$ and for each agent $i$, we can compute a point $\hat{x}_i^{(k)}$ satisfying the following inequality for any $z \in \mathcal{Z}$:
 \[
  (z - \hat{x}_i^{(k)})^T ( \nabla \theta_i(\hat{x}_i^{(k)}) - \gamma_i^{(k-1)} - \rho(\hat{x}_i^{(k)} - z^{(k-1)}) ) \le \xi.
 \]
 Suppose we modify \refline{alg:x-update}{alg:PPGA} by replacing $x_i^{(k)}$ with $\hat{x}_i^{(k)}$.
 This modification does not affect the asymptotic truthfulness guarantee of the algorithm.
 However, it slightly alters the algorithm's approximation of the core.
 In particular, it can be verified that modified versions of \reflemm{lemm:core-1} and \reflemm{lemm:residual} continue to hold, with additive error terms of $\xi$ and $2\sqrt{n\xi}/(\sqrt{\rho}K)$ on the right-hand sides of \refequ{eq:core-1} and \refequ{eq:residual}, respectively.

 Choosing $\xi = \Theta(1/\sqrt{n})$ ensures that \reflemm{lemm:approx-core}, and therefore \refthm{thm:asympt-core}, continue to hold for the modified algorithm.
 Thus, the modified algorithm preserves the asymptotic guarantees of the original, while achieving a total running time of $\mathcal{O}(n^{2.75} \log(n) m \log(m))$.
\end{proof}
\section{Experiments}
\label{sec:experiments}

In this section, we aim to show that \name can be deployed in practice to solve large-scale public-good allocation problems.
To this end, we implement \refalg{alg:PPGA} in Python using \texttt{CVXPY}, an open-source Python-embedded modeling language for convex optimization problems~\cite{diamond2016cvxpy}.
\name is highly parallelizable, particularly in the concurrent computation of $x$ and $\gamma$ for all agents.
We leverage this feature in our implementation by distributing the computational workload across multiple processes using Python's \texttt{multiprocessing} package.
The code for our implementation is provided at \github.

To conduct experiments, we leverage real-world data from \url{Pabulib.org}, an open participatory budgeting library~\cite{faliszewski2023participatory}.
Our experiments focus on 12 election instances, selected primarily based on the size of their voter population and the average number of approved projects per voter%
\footnote{We selected representative instances from about 60 instances that had at least 10k votes.}.
Each instance involves a collection of projects with associated costs and a designated total budget.
Voters express their preferences for the projects by casting approval votes for one or more projects.
We summarize the key characteristics of these election instances in \refapp{sec:instances}, and full details of each instance, such as project costs, are provided with our code (located in the \texttt{final\_data} folder).

As just mentioned, the instances involve approval votes and indivisible projects.
We utilized these instances to derive new ones wherein agents have cardinal utilities, and fractional allocations are deemed acceptable.
Fractional budget allocations are inspired by the motivating examples in the introduction and various related works~\cite{fain2016core,caragiannis2022truthful,freeman2019truthful,freeman2024project}.
We transform approval votes into cardinal utilities according to the \emph{cost-utility} approach~\cite{faliszewski2023participatory} using the following procedure:
For each voter $i$ and project $j$, we set $u_{ij} = 0$ if voter $i$ does not approve project $j$, and $u_{ij} = 1$ otherwise.
This ensures that voters' utilities are proportional to the budget allocated to the projects they support%
\footnote{Let $P_i$ be the set of projects supported by voter $i$.
Then, $i$'s utility is given by $U_i(z) = \sum_{j \in P_i} z_j$, where $z_j \leq s_j / c$ represents the fraction of the total budget allocated to project $j$.
This ensures that $i$'s utility is proportional to the budget allocated to the projects they support.}.

In the concluding remarks of \refsec{subsec:dp-for-mnw}, we provide guidelines for the DP parameters to guarantee our asymptotic properties.
There are also established practical norms for acceptable $\epsilon$ and $\delta$ values.
Following these norms, we set $\epsilon=c_{\epsilon}/\log(n)$, $\delta=c_{\delta}/\sqrt{n}$, and $K=c_{K}n$, where $c_{\epsilon}=1.5$, $c_{\delta}=0.3$, and $c_{K}=0.001$.
We further set $\alpha$ such that $\log(1/\delta)/(\alpha - 1) = \epsilon/2$.
This way, values for $\epsilon$ and $\delta$ approximate 0.3 and 0.001, respectively, keeping the noise magnitude, $\EE\left[\Vert q^{(k)} \Vert^2_2\right]$, under 3e-4 for the majority of instances.
We note that in our experiments, we set $\upsilon = 0$.
The introduction of $\upsilon$ as a parameter was solely motivated by a technical requirement to ensure that $\theta_i$ is a smooth function.
However, this smoothness condition has negligible practical significance.

We compare \name with the core%
\footnote{We find the core by solving the convex optimization of \reflemm{lemm:core-solution} using \refalg{alg:PPGA} without adding noise.}
using the following metrics:
\begin{itemize}[noitemsep=1pt,topsep=3pt,leftmargin=*]
	\item \textbf{Social welfare (SW)} for an allocation $z$ is defined as $\frac{1}{n}\sum_i U_{i}(z)$.
	SW serves as an indicator of the overall satisfaction achieved collectively by all agents from the allocation.
	\item \textbf{Proportionality score (PS)} of voter $i$ for an allocation $z$ is defined as the ratio of $i$'s utility for $z$ to $i$'s maximum attainable utility, i.e., $\frac{U_i(z)}{\max_{z\pr \in \mathcal{Z}}U_i(z\pr)}$.
	If the PS value is $\ge 1/n$ for all voters (or equivalently, if the minimum value of PS across voters multiplied by $n$ is $\ge 1$), then the allocation is \emph{proportional} ($\vert A \vert = 1$ in \refdef{def:core}).
	We report both the minimum (multiplied by $n$) and the average of PS values across all voters.
	\item \textbf{Statistical distance (SD)} between an allocation $z$ and a core solution $z^*$ is measured by their \emph{total variation distance}, defined as $\frac{1}{2}\Vert z - z^* \Vert_1$.
	Two allocations over $m$ items are considered statistically close if their total variation distance is a negligible function in $m$.
	To facilitate comparison, we normalize the total variation distance by dividing it by $m$.

\end{itemize}
For each metric, we report the average value over 50 runs.

\begin{table}[!t]
	\small
	\centering
	\begin{tabular}{c | c c | c c | c}
		\toprule
		\multirow{2}{*}{\textbf{Inst.}}& \multicolumn{2}{c|}{\textbf{Core's PS}} & \multicolumn{2}{c|}{\textbf{\name{}'s PS}} & \textbf{SD}\\
		& \textbf{Min} {\scriptsize$\bm{(\times n)}$} & \textbf{Avg} & \textbf{Min} {\scriptsize$\bm{(\times n)}$} & \textbf{Avg} & {\scriptsize$\bm{(\div m)}$} \\
		\midrule[\heavyrulewidth]
		1	& 90.7	& 0.27	& 111.9		& 0.27	& 0.00007	\\
		2	& 236.4	& 0.30	& 17.1		& 0.29	& 0.00016	\\
		3	& 235.5	& 0.18	& 191.1		& 0.18	& 0.00014	\\
		4	& 216.1	& 0.39	& 37.5		& 0.38	& 0.00023	\\
		5	& 15.0	& 0.33	& 14.3		& 0.33	& 0.00010	\\
		6	& 244.7	& 0.39	& 39.9		& 0.38	& 0.00030	\\
		7	& 11.0	& 0.29	& 11.1		& 0.29	& 0.00045	\\
		8	& 122.6	& 0.33	& 128.3		& 0.32	& 0.00008	\\
		9	& 163.5	& 0.34	& 168.0		& 0.34	& 0.00002	\\
		10	& 154.4	& 0.16	& 106.9		& 0.16	& 0.00034	\\
		11	& 519.8	& 0.45	& 513.4		& 0.45	& 0.00002	\\
		12	& 261.3	& 0.57	& 130.0		& 0.57	& 0.00003	\\
		\bottomrule
	\end{tabular}
	\caption{Proportionality score and statistical distance.}
	\label{tab:core-vs-PPGA}
\end{table}

\begin{figure}[!t]
	\pgfplotstableread[col sep=comma]{
x, core, mean, ci
1, 1, 0.995, 0.00005
2, 1, 0.980, 0.00048
3, 1, 0.989, 0.00009
4, 1, 0.976, 0.00040
5, 1, 0.999, 0.00009
6, 1, 0.974, 0.00061
7, 1, 0.991, 0.00014
8, 1, 0.995, 0.00004
9, 1, 0.998, 0.00003
10, 1, 0.985, 0.00016
11, 1, 0.997, 0.00003
12, 1, 0.997, 0.00004
}\performancedata

\begin{tikzpicture}[font=\scriptsize]
    \begin{axis}[
    	set layers=axis lines on top,
		ybar,
		bar width=.28cm,
		width=0.9\linewidth,
		height=4.2cm,
		ylabel={Normalized SW},
		xlabel={Election instances},
		grid,
		symbolic x coords={1, 2, 3, 4, 5, 6, 
                        7, 8, 9, 10, 11, 12},
		xtick=data,
		legend style={
		at={(0.9,1)},
		anchor=north, 
		legend columns=4,
		fill=none,
    	/tikz/every even column/.append style={column sep=4mm},
    	draw=none,
		},
		ymin=0.94,ymax=1.02,
		ytick={0.94,0.96,0.98,1.00,1.02},
		yticklabels={0.94,0.96,0.98,1.00,1.02},
		ylabel near ticks,
		xlabel near ticks
    ]
		\addplot+[style={draw=none,fill=mine}, error bars/.cd, y dir=both, y explicit, error bar style=black] table[x=x, y=mean, y error=ci]{\performancedata};
    \end{axis}
\end{tikzpicture}
	\caption{Social welfare of \name normalized to that of core (w/ 95\% confidence band).}
	\label{fig:sw}
\end{figure}

\reffig{fig:sw} illustrates the social welfare under \name normalized to that under the core solution, while \reftab{tab:core-vs-PPGA} summarizes proportionality scores and statistical distances across all election instances.
These results uncover several crucial insights.
Firstly, the statistical distance between the budget allocation under \name and the core solution remains consistently close to zero in all instances, hovering below 0.00045 for all cases.
Secondly, the observed discrepancy in social welfare values between \name and the core solution consistently falls below 3\% across all election instances.
Lastly, the minimum PS value $\times n$ exceeds 1 for all instance, indicating that \name satisfies the proportionality criteria for all instances.
The average PS values tend to be slightly higher under the core solution for some instances, but the discrepancy between the average PS values under \name and the core remains below 4\% in all instances.
Collectively, these findings strongly signify the high level of fairness achieved by \name.

Our empirical findings not only corroborate the theoretical results in \refsec{sec:analysis} but also illustrate that \name{} yields solutions that are statistically close to the core solution for all election instances.
We expect this result to hold for any instance with a large population and a linear utility model.
This expectation is based on our proof in \refthm{thm:asympt-core}, wherein we demonstrate that the distance between $\bar{x}$ and $\hat{z}$ asymptotically approaches zero with high probability.
For any linear utility model, it can be shown that the distance between $\bar{x}$ and $z^*$ also asymptotically approaches zero with high probability.
Consequently, the statistical distance between $\hat{z}$ and $z^*$ is asymptotically negligible for any linear utility model.
For other concave utility models, the statistical distance between $\hat{z}$ and $z^*$ might be higher, depending on the curvature of the function.
Nevertheless, one can demonstrate that the difference in the value of the NW objective for $\hat{z}$ and $z^*$ asymptotically approaches zero with high probability, implying similar results for PS.
\section{Related Works}\label{sec:related-works}

\textbf{Fair resource allocation}
without money (also known as cake cutting) has been extensively studied in the literature for private goods~\cite{procaccia2013cake}.
For public goods, the fair allocation problem has been studied in various contexts, including fair public decision-making~\cite{conitzer2017fair}, multi-agent knapsack problems~\cite{fluschnik2019fair}, multi-winner elections~\cite{munagala2022approximate}, and participatory budgeting~\cite{peters2021proportional}.
The truthful aggregation of agents' preferences has also been explored in public decision-making~\cite{procaccia2009approximate,goel2019knapsack,freeman2019truthful,caragiannis2022truthful,freeman2024project}.
However, the settings in these works differ from ours, as they aim to maximize social welfare and focus on $\ell_1$ preferences%
\footnote{An agent's disutility for an allocation is equal to the $\ell_1$ distance between that allocation and the agent's most preferred allocation.},
whereas our focus is on concave preferences and maximizing Nash welfare.


The work most closely related to this paper is that of Fain et al.~\cite{fain2016core}%
\footnote{Their notion of the core is based on capacity, where a blocking coalition receives a proportional share of the capacity rather than a proportional share of utility (see \refdef{def:core}).},
which finds an approximate core solution with high probability while achieving approximate truthfulness.
However, due to its reliance on several approximations, their approach fails to produce an asymptotic core solution.
As the number of agents increases, the approximation error for fairness (core) may grow.
In contrast, our approximation guarantee does not suffer from this issue.
By combining the Gaussian mechanism with ADMM to directly optimize the NW objective, our method ensures asymptotic truthfulness and finds an asymptotic core solution with high probability.


\textbf{Differentially private convex programming}
has been utilized in recent years to allocate private goods~\cite{hsu2014private,cummings2015privacy,hsu2016jointly,huang2018near,huang2019scalable}.
These methods often employ the dual ascent technique as a key tool~\cite{boyd2011distributed}.
The dual ascent method involves a sequence of two updates: the primal update, which optimizes the Lagrangian while fixing the dual variable, and the dual update, which takes a gradient ascent step to update the dual variable given the optimized primal variable.
However, the dual ascent method cannot be used for maximizing the NW objective, because, as we show in \refsec{sec:algorithm}, the Lagrangian for the convex program is an affine function of some components of the primal variable.
This causes the primal update to fail, as the dual problem is unbounded below for most values of the dual variable~\cite{boyd2011distributed}.
We avoid this by optimizing the augmented Lagrangian instead of the Lagrangian.

\textbf{Differentially private ADMM}
methods have also been extensively studied~\cite{shi2014linear,huang2015differentially,zhang2016dynamic,zhang2018improving,huang2019dp,iyengar2019towards}.
Although related, our work differentiates itself from these works in several aspects.
Firstly, while previous studies focus on the convergence rate of the objective function, we study the convergence of a primal variable to an approximate core solution.
To the best of our knowledge, our work is first to prove an asymptotic, game-theoretic property for a primal variable within differentially private ADMM\@.
Secondly, unlike prior work that introduces noise to the local variables, \name adds noise to the global variable (as detailed in \refsec{sec:algorithm}).
Finally, many studies on differentially private ADMM rely on a restrictive assumption regarding the strong convexity of the objective function, which does not hold for the NW objective.
\section{Conclusion}\label{sec:conclusion}
In this paper, we introduce \name, a mechanism designed for the fair and truthful allocation of divisible public goods.
\name achieves fairness by directly maximizing the NW objective and ensures truthfulness by deploying the Gaussian mechanism from differential privacy.
We showed that \name is asymptotically truthful and finds an asymptotic core solution with high probability.
By conducting experiments using real-world data from participatory budgeting elections, we showcased the practical applicability of \name.

\bibliographystyle{ACM-Reference-Format}
\bibliography{main}

\newpage
\appendix
\section{Notations}
\label{sec:notations}

\begin{table}[!ht]
	\small
	\renewcommand\thetable{}
	\centering
	\begin{tabular}{ cl }
	    \toprule
	    \textbf{Notation} & \textbf{Description} \\
	    \toprule
	    $n$ & Number of agents \\
	    $m$ & Number of public items \\
	    $s_j$ & Size of item $j$ \\
	    $s$ & Size vector, i.e., $(s_1,\dots, s_m)$\\
	    $c$ & Total capacity \\
		$z_j$ & Fraction of the total capacity that is allocated to item $j$ \\
	    $z$ & Allocation variable, i.e., $(z_1, \dots, z_m)$\\
		$\mathcal{Z}$ & Set of all feasible allocations, i.e., $\{z \in [0, 1]^m \given \Vert z \Vert_1 \le 1, \; cz \le s\}$ \\
	    $U_i(z)$ & Agent $i$'s utility function for allocation $z$\\
	    $u_i$ & Agent $i$'s utility vector, i.e., parameters of $U_i$: $(u_{i1},\dots,u_{id})$ \\
	    $\mathcal{U}$ & Set $[0,1]^d$ \\
	    $u$ & Utility vectors for all agents, i.e., $(u_1, \dots, u_n)$\\
	    $u_{-i}$ & Utility vector of all agents except agent $i$, i.e., $(u_1, \dots, u_{i-1}, u_{i+1}, \dots, u_n)$\\
	    $M(u)$ & Randomized mechanism that maps $u \in \mathcal{U}^n$ to probability distribution over $\mathcal{Z}$ \\
	    $x_i$ & Feasible allocation of agent $i$ \\
		$x$ & Vector of allocations, i.e., $(x_1, \dots, x_n) \in \mathcal{Z}^n$\\
		$\theta_i(x_i)$ & Smoothed logarithm of agent $i$'s utility, i.e., $\log(U_i(x_i) + \epsilon)$\\
	    $\theta(x)$ & Summation of $\theta_i$'s: $\sum_i \theta_i(x_i)$\\
	    $L$ & Lipschitz parameter of $U_i(z)$'s \\
	    $\epsilon$ & Multiplicative approximation factor for truthfulness, core, and DP \\
	    $\delta$ & Additive approximation factor for truthfulness, core, and DP \\
	    $\alpha$ & R{\'e}nyi divergence parameter \\
	    $\mathcal{N}(\mu,\Sigma)$ & Multivariate normal distribution with mean vector $\mu$ and covariance matrix $\Sigma$ \\
	    $K$ & Total number of iterations in \refalg{alg:PPGA} \\ 
	    $z^{(k)}$ & Global allocation variable at iteration $k$, i.e., $(z_{i1}^{(k)}, \dots, z_{im}^{(k)})$ \\
	    $x_i^{(k)}$ & Agent $i$'s local allocation variable at iteration $k$, i.e., $(x_{i1}^{(k)}, \dots, x_{im}^{(k)})$ \\
	    $x^{(k)}$ & Vector of local allocations at iteration $k$, i.e., $(x_1^{(k)}, \dots, x_n^{(k)})$ \\
	    $\gamma_i^{(k)}$ & Dual variable for $z = x_i$ constraint at iteration $k$, i.e., $(\gamma_{i1}^{(k)}, \dots, \gamma_{im}^{(k)})$ \\
	    $\gamma^{(k)}$ & Vector of dual variables, i.e., $(\gamma_1^{(k)}, \dots, \gamma_n^{(k)})$\\
	    $q^{(k)}$ & Multivariate Gaussian noise added to $z^{(k)}$ at iteration $k$ \\
	    $\sigma^2$ & Variance of added noise to each dimension of $z$ \\
	    $\rho$ & Penalty parameter for the augmented Lagrangian \\
	    $L_i^\rho$ & Agent $i$'s partial augmented Lagrangian with parameter $\rho$ \\
		$L^\rho$ & Summation of partial augmented Lagrangian functions, i.e., $sum_i L_i^\rho$ \\
		$\eta$ & Regularization parameter for the linearized augmented Lagrangian \\
		$L_i^{\rho,\eta}$ & Agent $i$'s linearized partial augmented Lagrangian with parameters $\rho$ and $\eta$ \\
	    $\Pi_\mathcal{Z}(z)$ & Euclidean projection of $z$ onto $\mathcal{Z}$, i.e., $\argmin_{z\pr\in \mathcal{Z}}\Vert z - z\pr \Vert_2^2$ \\
	    $\bar{z}$ & Time average of $z^{(k)}$'s, i.e., $(1/K)\sum_{k = 1}^{K}z^{(k)}$ \\
	    $\hat{z}$ & Euclidean projection of $\bar{z}$ onto $\mathcal{Z}$, i.e., $\Pi(\bar{z})$ \\
	    \bottomrule
	\end{tabular}
	\caption{List of notations}
	\label{tab:table}
\end{table}

\newpage

\section{Election Instances}
\label{sec:instances}

\begin{table}[!ht]
	\small
	\centering
	\begin{tabular}{c l c c c c}
		\toprule
		\multirow{2}{*}{\textbf{Inst.}} & \multirow{2}{*}{\textbf{Election}} & \textbf{\# Voters} & \textbf{\# Proj.} & \textbf{Budget} & \textbf{Avg. \# votes} \\
		 &  & $\bm{(n)}$ & $\bm{(m)}$ & $\bm{(c)}$ & \textbf{per voter} \\
		\midrule[\heavyrulewidth]
		\midrule[\heavyrulewidth]
		1 &	    Wroclaw'17 &					62,529 &	50 &	4,000,000 &	    1.8	\\
		2 &	    Warszawa'20 Praga Poludnie &	14,897 &	134 &	5,900,907 &	    9.1	\\
		3 &	    Katowice'21 &					36,370 &	47 &	3,003,438 &	    1.5	\\
		4 &	    Warszawa'21 Mokotow &			12,933 &	98 &	7,147,577 &	    9.7	\\
		5 &	    Wroclaw'16 Rejon NR 10-750 &	12,664 &	13 &	750,000 &       1	\\
		6 &	    Warszawa'23 Mokotow &			11,067 &	81 &	8,697,250 &	    9.1	\\
		7 &	    Wroclaw'16 Rejon NR 12-250 &	10,711 &	15 &	650,000 &	    1	\\
		8 &	    Wroclaw'16 &					67,103 &	52 &	4,500,000 &	    1.8	\\
		9 &	    Warszawa'22 &					81,234 &	129 &	28,072,528 &	7.9	\\
		10 &	Gdansk'20 &					    30,237 &	28 &	3,600,000 &	    1	\\
		11 &	Warszawa'21 &					95,899 &	106 &	24,933,409 &	8.3	\\
		12 &	Warszawa'20 &					86,721 &	101 &	24,933,409 &	7.2	\\
		\bottomrule
	\end{tabular}
	\caption{Characteristics of election instances.}
	\label{tab:table2}
\end{table}
\section{Supplementary Claims}
\label{sec:supp}

\begin{claim}
    \label{clm:receprocal}
    Let $f: \mathcal{D} \mapsto \RR$ be strictly positive and concave.
    Then, $F(x) = \frac{1}{f(x)}$ is convex.
\end{claim}

\begin{proof}
    First, not that since $f$ is strictly positive, $F$ is well-defined over $D$.
    To show that $F$ is convex, we need to verify that for any $x, x\pr \in D$ and any $\lambda \in [0,1]$:
    \[
        F(\lambda x + (1-\lambda)x\pr) \le \lambda F(x) + (1-\lambda)F(x\pr).
    \]
    Since $f$ is concave, it satisfies $f(\lambda x + (1-\lambda)x\pr) \ge \lambda f(x) + (1-\lambda)f(x\pr)$ for any $\lambda \ge 0$.
    Because $f$ is strictly positive, all terms are positive, and taking reciprocals reverses the inequality:
    \[
        \frac{1}{f(\lambda x + (1-\lambda)x\pr)} \le \frac{1}{\lambda f(x) + (1-\lambda)f(x\pr)} \le \lambda \frac{1}{f(x)} + (1-\lambda)\frac{1}{f(x\pr)},
    \]
    where the final inequality follows from Jensen's inequality applied to the convex function $t \mapsto 1/t$ on the positive reals.
    Thus, $F(x)$ is convex.
\end{proof}

\begin{claim}
    \label{clm:gradient}
    If $f: \mathcal{D} \mapsto \RR$ is concave and $L$-Lipschitz continuous, then $\Vert \nabla f \Vert_2 \le L$.
    Also, if $f$ is concave and $\Vert \nabla f \Vert_2 \le L$, then $f$ is $L$-Lipschitz continuous.
\end{claim}

\begin{proof}
    First, we show that if $f$ is concave and $L$-Lipschitz continuous, then $\Vert \nabla f \Vert_2 \le L$.
    Since $f$ is concave, for all $x, x\pr \in D$, we have:
    \[
        f(x\pr) \le f(x) + {\nabla f(x)}^T (x\pr - x).
    \]
    Set $x\pr = x - \delta \frac{\nabla f(x)}{\Vert \nabla f(x) \Vert_2}$ for some $0 < \delta < D$, where $D = \sup_{x,x\pr \in \mathcal{D}} \Vert x - x\pr \Vert_2$.
    Then, we have:
    \[
        \delta \Vert \nabla f(x) \Vert_2 \le f(x) - f(x\pr) \le L \Vert x - x\pr \Vert_2 = L \delta,
    \]
    where the last inequality is due to Lipschitz continuity of $f$.
    Dividing both sides by $\delta$ yields the desired result.

    Next, we show that if $f$ is concave and $\Vert \nabla f \Vert_2 \le L$, then $f$ is $L$-Lipschitz continuous.
    This follows directly from concavity and boundedness of the gradient:
    \[
        f(x\pr) - f(x) \le {\nabla f(x)}^T (x\pr - x) \le \Vert \nabla f(x) \Vert_2 \Vert x\pr - x \Vert_2 \le L \Vert x\pr - x \Vert_2.
    \]
    By switching $x$ and $x\pr$ in the above inequality, we can show:
    \[
        f(x) - f(x\pr) \le L \Vert x - x\pr \Vert_2,
    \]
    for the same $x$ and $x\pr$.
\end{proof}

\begin{claim}
    \label{clm:smooth2lips}
    Let $f: \mathcal{D} \mapsto \RR$ be concave, $\beta$-smooth, and bounded between 0 and 1.
    Suppose that $\mathcal{D}$ is a bounded convex set, such that $\sup_{x,x\pr \in \mathcal{D}} \Vert x - x\pr \Vert_2 \le D < \infty$.
    Then, $f$ is $L$-Lipschitz, with $L \le \frac{1}{D} + \frac{\beta D}{2}$.
\end{claim}

\begin{proof}
    Given that $f$ is concave and $\beta$-smooth, for all $x,x^\prime \in \mathcal{D}$, we have:
    \[
        f(x) + {\nabla f(x)}^T (x\pr - x) - f(x\pr) \le \frac{\beta}{2} \Vert x\pr - x \Vert_2^2.
    \]
    Setting $x^\prime = x + \delta \frac{\nabla f(x)}{\Vert f(x) \Vert_2}$, where $0 < \delta \le D$, we have:
    \[
        \delta \Vert \nabla f(x) \Vert_2 \le f(x\pr) - f(x) + \frac{\beta}{2} \delta^2 \le 1 + \frac{\beta}{2} \delta^2.
    \]
    Dividing by $\delta > 0$ and substituting $\delta = D$ gives us the desired upper bound on $\nabla f(x)$.
\end{proof}

\begin{claim}
    \label{clm:log-smooth}
    Let $f: \mathcal{D} \mapsto \RR$ be concave and bounded between 0 and 1.
    Suppose that $\mathcal{D}$ is a bounded convex set, such that $\sup_{x,x\pr \in \mathcal{D}} \Vert x - x\pr \Vert_2 \le D < \infty$.
    If $f$ is $\beta$-smooth, then for $\upsilon > 0$, $h(x) = \log(f(x) + \upsilon)$ has $L$-Lipschitz gradient (i.e., is smooth), with $L \le \frac{M^2}{\upsilon^2}  + \frac{\beta}{\upsilon}$, where $M = \frac{1}{D} + \frac{\beta D}{2}$.
\end{claim}

\begin{proof}
    For any $x, x\pr \in \mathcal{D}$, we have:
    \begin{align*}
    \Vert \nabla h(x) - \nabla h(x\pr) \Vert_2 &= \left\Vert \frac{\nabla f(x)}{f(x) + \upsilon} - \frac{\nabla f(x\pr)}{f(x\pr) + \upsilon} \right\Vert_2  \\
    &= \left\Vert \left( \frac{1}{f(x) + \upsilon} - \frac{1}{f(x\pr) + \upsilon} \right) \nabla f(x) + \frac{\nabla f(x) - \nabla f(x\pr)}{f(x\pr) + \upsilon} \right\Vert_2 \\
    &\le \left\vert \frac{1}{f(x) + \upsilon} - \frac{1}{f(x\pr) + \upsilon} \right\vert \Vert \nabla f(x) \Vert_2 + \frac{\Vert \nabla f(x) - \nabla f(x\pr) \Vert_2}{f(x\pr) + \upsilon} \\
    &= \frac{\vert f(x) - f(x\pr) \vert}{(f(x) + \upsilon) (f(x\pr) + \upsilon)} \Vert \nabla f(x) \Vert_2 + \frac{\Vert \nabla f(x) - \nabla f(x\pr) \Vert_2}{f(x\pr) + \upsilon} \\
    &\le \frac{M^2}{\upsilon^2}  \Vert x - x\pr \Vert_2 + \frac{\beta}{\upsilon}  \Vert x - x\pr \Vert_2 \\
    &= \left(\frac{M^2}{\upsilon^2}  + \frac{\beta}{\upsilon}\right) \Vert x - x\pr \Vert_2,
\end{align*}
    where the last inequality follows from $f$'s $M$-Lipschitz continuity (\refclm{clm:smooth2lips}) and $\beta$-smoothness.
\end{proof}
\section{Omitted Proofs}
\label{sec:app}

\subsection{Proof of \reflemm{lemm:core-solution}}
\label{subsec:proof-lemm-core-solution}
\begin{proof}
 By concavity of $U_i$, for all $z, z\pr \in \mathcal{Z}$, we have:
 \begin{equation}
 \label{eq:concave}
  U_i(z\pr) - U_i(z) \le \nabla U_i(z)^T (z\pr - z).
 \end{equation}
 Let $z^*$ be an MNW solution.
 The first-order optimality condition for $z^*$ requires that the following inequality holds for all $z\pr \in \mathcal{Z}$:
 \begin{equation}
 \label{eq:MNW}
  \sum_i \frac{\nabla U_i(z^*)^T}{U_i(z^*)} (z\pr - z^*) \le 0 \; \xRightarrow{\text{by }\refequ{eq:concave}} \;
  \frac{1}{n}\sum_i \frac{U_i(z\pr)}{U_i(z^*)} \le 1.
 \end{equation}
 For contradiction, suppose that $z^*$ is not a core outcome.
 Then, there exists a set of agents $A$ and an allocation $z^\prime$ such that $(\vert A \vert/n)U_i(z\pr) \ge U_i(z^*)$, and at least one inequality is tight.
 This implies $(1/n)\sum_{i\in A} U_i(z\pr)/U_i(z^*) > 1$, which contradicts \refequ{eq:MNW}.
\end{proof}

\subsection{Proof of \reflemm{lemm:mnw-core-approx}}
\label{subsec:proof-lemm-mnw-core-approx}
\begin{proof}
 Suppose, for contradiction, that $z$ is not a core solution.
 Then, there must exist a set $A$ and some $z^\prime \in \mathcal{Z}$ such that $(\vert A \vert / n)U_i(z\pr) \ge (1+\epsilon)U_i(z) + \delta$ for all $i \in A$, with at least one strict inequality.
 This implies: $(1/n)\sum_{i\in A} U_i(z\pr)/(U_i(z) + \delta/(1 + \epsilon)) > 1 + \epsilon$, contradicting \refequ{eq:sum-tilde-z}.
\end{proof}

\subsection{Proof of \reflemm{lemm:core-1}}
\label{subsec:proof-lemm-core-1}
To prove \reflemm{lemm:core-1}, we first present the following lemma, which relates $\tilde{w}^{(k)}$ to any $w \in \mathcal{W}_{\mathcal{Z}}$, where
$\mathcal{W}_{\mathcal{Z}} \triangleq \{ (x, z, \gamma) \in \mathcal{W} \given x = Gz \}$:
\begin{lemma}
\label{lemm:main-inequ}
 Let $\{w^{(k)}\}$ and $\{q^{(k)}\}$ be sequences produced by \refalg{alg:PPGA}.
 Then, the following inequality holds for any $w \in \mathcal{W}_\mathcal{Z}$:
 \begin{align}
 \label{eq:main-inequ}
  (x - &x^{(k)})^T \nabla\theta(x^{(k)}) + \gamma^{T}(x^{(k)} - Gz^{(k)}) \le \rho (z^{(k)} - z)^T q^{(k)} \\
  &+ \frac{n\rho}{2}\left( \Vert z - z^{(k-1)}\Vert_2^2 - \Vert z - z^{(k)}\Vert_2^2 \right) + \frac{1}{2\rho}\left( \Vert \gamma - \gamma^{(k-1)} \Vert_2^2 - \Vert \gamma - \gamma^{(k)}\Vert_2^2 \right).\nonumber
 \end{align}
\end{lemma}

\begin{proof}
 The first-order optimality conditions corresponding to the update step in \refline{alg:x-update}{alg:PPGA} imply the following inequality for all $i$ and $z \in \mathcal{Z}$.
 \begin{equation}
 \label{eq:opt-x}
  (z -  x_i^{(k)})^T ( \nabla\theta_i(x_i^{(k-1)}) - \gamma_i^{(k-1)} - \rho(x_i^{(k)} - z^{(k-1)}) ) \le 0.
 \end{equation}
 Let $\tilde{\gamma}^{(k)} \triangleq \gamma^{(k-1)} + \rho(x^{(k)} - G z^{(k-1)})$.
 Then, we can rewrite \refequ{eq:opt-x} as:
 \[
  (z -  x_i^{(k)})^T ( \nabla\theta_i(x_i^{(k)}) - \tilde{\gamma}_i^{(k)} ) \le 0.
 \]
 Summing this over all $i$, for any $z \in \mathcal{Z}$ and $x = Gz$, we have:
 \begin{equation}
 \label{eq:x-update1}
  (x -  x^{(k)})^T \nabla\theta(x^{(k)}) - (x -  x^{(k)})^T\tilde{\gamma}^{(k)} \le 0.
 \end{equation}
 Next, given \refequ{eq:def-noisy-admm2}, \refline{alg:z-update}{alg:PPGA} implies that $z^{(k)}$ is a solution to:
 \[
  \underset{z}{\text{maximize}} \quad \sum_i \left( -(\gamma_{i}^{(k-1)})^T (x_{i}^{(k)} - z + q^{(k)}) - \frac{\rho}{2} \Vert x_{i}^{(k)} - z + q^{(k)} \Vert_2^2 \right).
 \]
 The first-order optimality conditions for this optimization imply:
 \begin{equation}
 \label{eq:opt-z}
  (z - z^{(k)})^T  \left(\sum_i \left( \gamma_i^{(k-1)} + \rho (x_i^{(k)} - z^{(k)} + q^{(k)}) \right)\right) \le 0 \quad \text{for all } z \in \RR^m.
 \end{equation}
 Given the definition of $\tilde{\gamma}^{(k)}$, we can rewrite \refequ{eq:opt-z} for all $z \in \RR^m$ as:
 \begin{align}
 \label{eq:z-update1}
  &(z - z^{(k)})^T \left(\sum_i \tilde{\gamma}_i^{(k)} - n \rho (z^{(k)} - z^{(k-1)}) +  n \rho q^{(k)}\right) \le 0 \; \Rightarrow \nonumber \\
  &(z - z^{(k)})^T \sum_i \tilde{\gamma}_i^{(k)} \le n\rho (z - z^{(k)})^T(z^{(k)} - z^{(k-1)}) -  n \rho (z - z^{(k)})^T q^{(k)}.
 \end{align}
 Next, given \refline{alg:g-update}{alg:PPGA}, for all $\gamma \in \RR^{mn}$ we have:
 \begin{align}
 \label{eq:g-update1}
  &x^{(k)} - Gz^{(k)} = (\gamma^{(k)} - \gamma^{(k-1)})/\rho \; \Rightarrow \nonumber \\
  &(\gamma - \tilde{\gamma}^{(k)})^{T}(x^{(k)} - Gz^{(k)}) = (\gamma - \tilde{\gamma}^{(k)})^{T}(\gamma^{(k)} - \gamma^{(k-1)})/\rho.
 \end{align}
 To put everything together, we use the following identity:
 \[
  (x^{(k)} - Gz)^T\tilde{\gamma}^{(k)} + (z - z^{(k)})^T \sum_i \tilde{\gamma}_i^{(k)} + (\gamma - \tilde{\gamma}^{(k)})^{T}(x^{(k)} - Gz^{(k)}) = \gamma^{T}(x^{(k)} - Gz^{(k)}).
 \]
 With this, we can combine \refequs{eq:x-update1}{eq:g-update1} to get the following inequality for any $w = \mathcal{W}_{\mathcal{Z}}$:
 \begin{align}
 \label{eq:all-update2}
 (x -  x^{(k)})^T \nabla\theta(x^{(k)}) + &\gamma^{T}(x^{(k)} - Gz^{(k)}) \le n\rho (z^{(k)} - z)^T q^{(k)} \\
 &+ n\rho (z - z^{(k)})^T(z^{(k)} - z^{(k-1)}) + (\gamma - \tilde{\gamma}^{(k)})^T(\gamma^{(k)} - \gamma^{(k-1)})/\rho. \nonumber
 \end{align}
 Next, we focus on the right-hand side of \refequ{eq:all-update2}.
 Given the following identity:
 \[
  2(a - b)^T(c - d) = \Vert a - d \Vert_2^2 - \Vert a - c\Vert_2^2 + \Vert b - c \Vert_2^2 - \Vert b - d\Vert_2^2,
 \]
 we have:
 \begin{align}
  2(z - z^{(k)})^T(z^{(k)} - z^{(k-1)}) = \Vert z - z^{(k-1)}\Vert_2^2 - \Vert z - z^{(k)}\Vert_2^2 &- \Vert z^{(k)} - z^{(k-1)}\Vert_2^2, \label{eq:z-zk} \\
  2(\gamma - \tilde{\gamma}^{(k)})^T(\gamma^{(k)} - \gamma^{(k-1)}) = \Vert \gamma - \gamma^{(k-1)} \Vert_2^2 - \Vert \gamma - \gamma^{(k)}\Vert_2^2 &- \Vert \tilde{\gamma}^{(k)} - \gamma^{(k-1)} \Vert_2^2 \nonumber \\
  &+ \Vert \tilde{\gamma}^{(k)} - \gamma^{(k)} \Vert_2^2. \label{eq:g-gk}
 \end{align}
 Given the definition of $\tilde{\gamma}^{(k)}$ and \refline{alg:g-update}{alg:PPGA}, we have:
 \begin{align}
 \label{eq:gk-gk-1}
  \Vert \tilde{\gamma}^{(k)} - \gamma^{(k)} \Vert_2^2 &= \Vert \rho(x^{(k)} - G z^{(k-1)}) - (\gamma^{(k)} - \gamma^{(k-1)}) \Vert_2^2 \nonumber \\
  &= \rho^2 \Vert x^{(k)} - G z^{(k-1)} - x^{(k)} + G z^{(k)} \Vert_2^2 \nonumber \\
  &= n\rho^2 \Vert z^{(k)} - z^{(k - 1)} \Vert_2^2.
 \end{align}
 Substituting \refequs{eq:z-zk}{eq:gk-gk-1} into \refequ{eq:all-update2} gives \refequ{eq:main-inequ}.
\end{proof}

We are now ready to prove \reflemm{lemm:core-1}:
\begin{proof}
 We start by rewriting $(x -  x^{(k-1)})^T \nabla\theta(x^{(k-1)})$ as:
 \[
  (x -  x^{(k)})^T \nabla\theta(x^{(k)}) = \sum_i \frac{(x_i - x_i^{(k)})^T \nabla U_i(x_i^{(k)})}{U_i(x_i^{(k)}) + \upsilon}.
 \]
 Since $U_i(x)$ is concave, for any $i$ and for any $x, x\pr \in \mathcal{Z}$, we have:
 \begin{equation}
 \label{eq:u_concave}
  U_i(x\pr) - U_i(x) \le (x\pr - x)^T\nabla U_i(x).
 \end{equation}
 Therefore, \refequ{eq:main-inequ} implies:
 \begin{align}
 \label{eq:main-inequ-2}
  \sum_i &\frac{U_i(x_i) + \upsilon}{U_i(x_i^{(k)})+ \upsilon} + \gamma^{T}(x^{(k)} - Gz^{(k)}) \le n + n\rho (z^{(k)} - z)^T q^{(k)} \\
  &+ \frac{n\rho}{2}\left( \Vert z - z^{(k-1)}\Vert_2^2 - \Vert z - z^{(k)}\Vert_2^2 \right) + \frac{1}{2\rho}\left( \Vert \gamma - \gamma^{(k-1)} \Vert_2^2 - \Vert \gamma - \gamma^{(k)}\Vert_2^2 \right). \nonumber
\end{align}
 Next, since \refequ{eq:main-inequ-2} holds for any $w \in \mathcal{W}_\mathcal{Z}$, it in particular holds when $\gamma = \bm{0}_{mn}$, which yields:
 \begin{align}
  \label{eq:main-inequ-xx}
  \frac{1}{n}\sum_i &\frac{U_i(z) + \upsilon}{U_i(x_i^{(k-1)}) + \upsilon} \le 1 + \rho (z^{(k)} - z)^T q^{(k)} \\
  &+ \frac{\rho}{2}\left( \Vert z - z^{(k-1)}\Vert_2^2 - \Vert z - z^{(k)}\Vert_2^2 \right) + \frac{1}{2n\rho}\left( \Vert \gamma^{(k-1)} \Vert_2^2 - \Vert \gamma^{(k)}\Vert_2^2 \right). \nonumber
 \end{align}
 For any $z \in \mathcal{Z}$, we have $\Vert z \Vert_2^2 \le \Vert z \Vert_1^2 \le 1$.
 Given this inequality, by summing \refequ{eq:main-inequ-xx} over $k = 1$ to $K$ and dividing by $K$, we obtain the following for any $z \in \mathcal{Z}$:
 \begin{equation}
 \label{eq:main-inequ-3}
  \frac{1}{n}\sum_i \frac{1}{K}\sum_{k=1}^{K} \frac{U_i(z) + \upsilon}{U_i(x_i^{(k)}) + \upsilon} \le 1 + \frac{\rho}{K} \sum_{k=1}^{K} (z^{(k)} - z)^T q^{(k)} + \frac{\rho}{2K}.
 \end{equation}
 Since $U_i(x_i) + \upsilon$ is strictly positive and concave, the function $1 / (U_i(x_i) \upsilon)$ is convex (\refclm{clm:receprocal}).
 As a result, by Jensen's inequality, it follows that for any $z \in \mathcal{Z}$, we have:
 \[
  \frac{1}{K}\sum_{k=1}^{K} \frac{U_i(z) + \upsilon}{U_i(x_i^{(k)}) + \upsilon} \ge \frac{U_i(z) + \upsilon}{U_i\left(\frac{1}{K}\sum_{k=1}^{K} x_i^{(k)}\right) + \upsilon} = \frac{U_i(z) + \upsilon}{U_i(\bar{x}) + \upsilon} \ge \frac{U_i(z)}{U_i(\bar{x}) + \upsilon}.
 \]
 Given the last inequality, \refequ{eq:main-inequ-3} implies \refequ{eq:core-1}.
\end{proof}

\subsection{Proof of \reflemm{lemm:residual}} 
\label{subsec:proof-lemm-residual}
\begin{proof}
 First, we note that, since the objective function of \refequ{eq:modified-opt} is continuous on a compact set, the problem attains its bounded global maximum.
 Therefore, there exist optimal solutions $z^*\in \mathcal{Z}$ and $x^* \in \mathcal{Z}^n$ such that $x^*_i = z^*$ for all $i$, which achieve this maximum value~\cite[Theorem 4.16]{rudin1976principles}.

 \noindent
 Second, \emph{strong duality} holds for \refequ{eq:modified-opt}.
 This follows from three facts: (i) the objective function is concave, (ii) the constraints are affine, and (iii) \emph{Slater's condition} is satisfied, i.e., there exists a \emph{strictly feasible} point that lies in the relative interior of $\mathcal{Z}$ and satisfies all constraints (e.g., $x_i = z = s / (c \Vert s \Vert)$ for all $i$).
 As a result, the dual optimal value is bounded and attained, and there exist optimal Lagrange multipliers $\gamma^*$ that achieves this value~\cite{boyd2004convex}.

 \noindent
 Next, Since $z^*$ is a solution to \refequ{eq:modified-opt}, the first-order optimality conditions require $\sum_i \gamma_i^* = 0$.
 Therefore, by setting $w = w^*$ in \refequ{eq:main-inequ}, we have:
 \begin{align}
  \label{eq:main-inequ-4}
  (x^* -  &x^{(k)})^T \nabla\theta(x^{(k)}) + {\gamma^*}^{T} x^{(k)} \le n\rho (z^{(k)} - z)^T q^{(k)} \\
  &+ \frac{n\rho}{2}\left( \Vert z^* - z^{(k-1)}\Vert_2^2 - \Vert z^* - z^{(k)}\Vert_2^2 \right) + \frac{1}{2\rho}\left( \Vert \gamma^* - \gamma^{(k-1)} \Vert_2^2 - \Vert \gamma^* - \gamma^{(k)}\Vert_2^2 \right).\nonumber
 \end{align}
 Since $x^*$ is a solution to \refequ{eq:modified-opt}, the first-order optimality conditions require:
 \begin{equation}
 \label{eq:opt-w-star}
  (x^{(k)} - x^*)^T(\nabla\theta(x^*) - \gamma^*) \le 0.
 \end{equation}
 Therefore, by summing \refequ{eq:main-inequ-4} and \refequ{eq:opt-w-star}, we obtain:
 \begin{align}
 \label{eq:main-inequ-5}
  (x^* - &x^{(k)})^T(\nabla\theta(x^{(k)}) - \nabla\theta(x^*)) \le n\rho (z^{(k)} - z^*)^T q^{(k)} \nonumber \\
  &+ \frac{n\rho}{2}\left( \Vert z^* - z^{(k-1)}\Vert_2^2 - \Vert z^* - z^{(k)}\Vert_2^2 \right) + \frac{1}{2\rho}\left( \Vert \gamma^* - \gamma^{(k-1)} \Vert_2^2 - \Vert \gamma^* - \gamma^{(k)}\Vert_2^2 \right).
 \end{align}
 Since $\theta(x)$ is concave, we have $(x - x\pr)^T(\nabla\theta(x) - \nabla\theta(x\pr)) \le 0$.
 Therefore, \refequ{eq:main-inequ-5} implies:
 \begin{equation*}
  \Vert \gamma^{(k)} - \gamma^* \Vert_2^2 - \Vert \gamma^{(k-1)} - \gamma^* \Vert_2^2 \le n\rho^2 \left( 2(z^{(k)} - z^*)^T q^{(k)} + \Vert z^* - z^{(k-1)}\Vert_2^2 - \Vert z^* - z^{(k)}\Vert_2^2 \right).
 \end{equation*}
 Given that $\Vert z^* \Vert_2^2 \le 1$, by summing this last inequality over $k = 1$ to $K$, we get:
 \begin{align}
 \label{eq:cont-2}
  \Vert \gamma^{(K)} - \gamma^* \Vert_2^2 \le 2n\rho^2 \sum_{k=1}^{K} (z^{(k)} - z^*)^T q^{(k)} + n\rho^2 + \Vert \gamma^* \Vert_2^2.
 \end{align}
 Next, we have:
 \begin{align*}
  \Vert \rho K (\bar{x} - G\bar{z}) \Vert_2^2 = \Vert \gamma^{(K)} \Vert_2^2 &= \Vert \gamma^{(K)} - \gamma^* + \gamma^* \Vert_2^2 \nonumber \\
  &\le 2 \Vert \gamma^* \Vert_2^2 + 2 \Vert \gamma^{(K)} - \gamma^* \Vert_2^2 \nonumber \\
  &\le 4n\rho^2 \sum_{k=1}^{K} (z^{(k)} - z^*)^T q^{(k)} + 2n\rho^2 + 4\Vert \gamma^* \Vert_2^2,
 \end{align*}
 which implies:
 \begin{equation}
  \label{eq:residual-2}
  \Vert \bar{x} - G\bar{z} \Vert_2 \le \frac{2\sqrt{n}}{K}\left(\sum_{k=1}^{K} \vert (z^{(k)} - z^*)^T q^{(k)} \vert \right)^{\frac{1}{2}} + \frac{\sqrt{2n}}{K} + \frac{2}{\rho K}\Vert \gamma^* \Vert_2.
 \end{equation}

 \noindent
 The Euclidean projection onto $\mathcal{Z}$ is contractive.
 Therefore, since $\bar{x}, \hat{z} \in \mathcal{Z}$, we have:
 \[
  \Vert \bar{x} - G\hat{z} \Vert_2 = \Vert \bar{x} - G\Pi_\mathcal{Z}(\bar{z}) \Vert_2 \le \Vert \bar{x} - G\bar{z} \Vert_2.
 \]
 Given this inequality, \refequ{eq:residual-2} implies \refequ{eq:residual}.
\end{proof}

\subsection{Proof of \reflemm{lemm:zkqk}}
\label{subsec:proof-lemm-zkqk}
\begin{proof}
 For any $z \in \mathcal{Z}$, we have:
 \begin{align}\label{eq:tail1}
  \vert (z^{(k)} - z)^T q^{(k)} \vert &= \vert (\frac{1}{n}\sum_i x^{(k)}_i + q^{(k)} - q^{(k-1)} - z)^T q^{(k)} \vert \nonumber \\
  &\le \vert (\frac{1}{n}\sum_i x^{(k)}_i)^T q^{(k)} \vert + \Vert q^{(k)} \Vert_2^2 + \vert {q^{(k-1)}}^T q^{(k)} \vert + \vert z^T q^{(k)} \vert \nonumber \\
  &\le \Vert \frac{1}{n}\sum_i x^{(k)}_i \Vert_1 \Vert q^{(k)} \Vert_2 + \Vert q^{(k)} \Vert_2^2 + \vert {q^{(k-1)}}^T q^{(k)} \vert + \Vert z \Vert_1 \Vert q^{(k)} \Vert_2 \nonumber \\
  &\le 2 \Vert q^{(k)} \Vert_2 + \Vert q^{(k)} \Vert_2^2 + \vert {q^{(k-1)}}^T q^{(k)} \vert.
 \end{align}
 Here, the first inequality follows from the triangle inequality.
 The second inequality follows from the Cauchy–Schwarz inequality and the fact that $\Vert \cdot \Vert_2 \le \Vert \cdot \Vert_1$--that is, for any vectors $a$ and $b$ in an inner product space, $\vert a^T b \vert \le \Vert a \Vert_2 \Vert b \Vert_2 \le \Vert a \Vert_1 \Vert b \Vert_2$.
 The third inequality holds because $\frac{1}{n}\sum_i x^{(k)}_i \in \mathcal{Z}$, and for any $z \in \mathcal{Z}$, we have $\Vert z \Vert_2 \le \Vert z \Vert_1 \le 1$.
 Finally, for the last term in \refequ{eq:tail1}, we apply Young's inequality to obtain:
 \begin{align*}
  \vert {q^{(k-1)}}^T q^{(k)} \vert  \le \frac{1}{2} \Vert q^{(k-1)} \Vert_2^2 + \frac{1}{2} \Vert q^{(k)} \Vert_2^2,
 \end{align*}
 Substituting the last inequality into \refequ{eq:tail1} and summing over $k$, we have:
 \begin{align}\label{eq:tail2}
  \vert\sum_{k = 1}^{K} (z^{(k)} - z)^T q^{(k)} \vert &\le \sum_{k = 1}^{K} \vert (z^{(k)} - z)^T q^{(k)} \vert \nonumber \\
  &\le 2 \sum_{k = 1}^{K} \left( \Vert q^{(k)} \Vert_2^2 + \Vert q^{(k)} \Vert_2 \right).
 \end{align}

 \noindent
 We next focus on tail behavior of $\Vert q^{(k)} \Vert_2^2$ and $\Vert q^{(k)} \Vert_2$ separately.
 Starting with $\Vert q^{(k)} \Vert_2^2$, note that each $q^{(k)}_j \sim \mathcal{N}(0,\sigma^2)$ is a \emph{sub-Gaussian} random variable%
 \footnote{A real-valued random variable $X$ is called sub-Gaussian if there exists a constant $\sigma > 0$ such that for all $t \in \RR$, $\EE [ \exp(t (X - \EE[X])) ] \le \exp((t^2 \sigma^2)/2)$.}.
 Therefore, by~\cite[Lemma 2.7.6]{vershynin2018high}, each $(q^{(k)}_j)^2$ is \emph{sub-exponential}%
 \footnote{A real-valued random variable $X$ is called sub-exponential if there exist constants $\nu, \alpha > 0$ such that for all $\vert t \vert < 1/\alpha$, $\EE[ \exp(t (X - \EE[X])) ] \le \exp((t^2 \nu^2)/2)$.}, with $\EE[(q^{(k)}_j)^2] = \sigma^2$ and
 \[
  \Vert (q^{(k)}_j)^2 - \sigma^2 \Vert_{\psi_1} \le C_1 \sigma^2,
 \]
 where $C_1$ is a constant, and $\Vert X \Vert_{\psi_1} = \inf \{t > 0 \given \EE[\exp(\vert X \vert / t)] \le 2\}$ denotes the \emph{sub-exponential norm} of a real-valued random variable $X$.
 Since $q^{(k)}_j$'s are i.i.d.\ across all $k$ and $j$, for any $t \ge 0$ and some constant $c_1$, \emph{Bernstein's inequality}~\cite[Theorem 2.8.1]{vershynin2018high} implies:
 \begin{equation}\label{eq:tail3}
  \PP\left[ \sum_{k=1}^{K} \Vert q^{(k)} \Vert_2^2 - Km\sigma^2 \ge t \right] \le \exp\left(-c_1\min\left(\frac{t^2}{K m \sigma^4}, \frac{t}{\sigma^2}\right)\right).
 \end{equation}

 \noindent
 Next, by~\cite[Theorem 3.1.1 and Lemma 2.6.8]{vershynin2018high}, $\Vert q^{(k)} \Vert_2$ is a sub-Gaussian random variable with
 \[
  \left\Vert \Vert q^{(k)} \Vert_2 - \EE\left[ \Vert q^{(k)} \Vert_2 \right] \right\Vert_{\psi_2} \le C_2\sigma^2,
 \]
 where $C_2$ is a constant, and $\Vert X \Vert_{\psi_2} = \inf \{t > 0 \given \EE[\exp(X^2 / t^2)] \le 2\}$ denotes the \emph{sub-Gaussian norm} of a real-valued random variable $X$.
 Since $q^{(k)}$'s are independent, by the \emph{general Hoeffding's inequality}~\cite[Theorem 2.6.2]{vershynin2018high}, for any $t \ge 0$ and some constant $c_2$, we have:
 \[
  \PP\left[ \sum_{k=1}^{K} \left( \Vert q^{(k)} \Vert_2 - \EE\left[ \Vert q^{(k)} \Vert_2 \right] \right) \ge t \right] \le \exp(-\frac{c_2 t^2}{K \sigma^4}).
 \]
 We next provide an upper bound on $\EE\left[ \Vert q^{(k)} \Vert_2 \right]$.
 Consider the inequality $\sqrt{u} \le (1 + u)/2$ which holds for any $u \ge 0$.
 By setting $u = \frac{1}{m\sigma^2}\Vert q^{(k)} \Vert_2^2$, we get:
 \[
  \frac{\Vert q^{(k)} \Vert_2}{\sqrt{m}\sigma} \le \frac{1 + (1/m\sigma^2) \Vert q^{(k)} \Vert_2^2}{2}.
 \]
 Taking expectations on both sides of the inequality, we obtain:
 \[
  \EE\left[ \Vert q^{(k)} \Vert_2 \right] \le \sqrt{m}\sigma ~ \frac{1 + 1}{2} = \sqrt{m}\sigma.
 \]
 Therefore, we have:
 \begin{equation}\label{eq:tail4}
  \PP\left[ \sum_{k=1}^{K} \Vert q^{(k)} \Vert_2 - K\sqrt{m}\sigma \ge t \right] \le \PP\left[ \sum_{k=1}^{K} \left( \Vert q^{(k)} \Vert_2 - \EE\left[ \Vert q^{(k)} \Vert_2 \right] \right) \ge t \right] \le \exp(-\frac{c_2 t^2}{K \sigma^4}).
 \end{equation}

 \noindent
 Given \refequs{eq:tail2}{eq:tail4} and the union bound, for $t\pr = 4t + 2Km\sigma^2 + 2K\sqrt{m}\sigma$, we have:
 \begin{align*}
  \PP\left[ \vert\sum_{k = 1}^{K} (z^{(k)} - z)^T q^{(k)} \vert \ge t\pr \right] &\le \PP\left[ 2 \sum_{k = 1}^{K} \left( \Vert q^{(k)} \Vert_2^2 + \Vert q^{(k)} \Vert_2 \right) \ge t\pr \right] \\
  &\le \PP\left[ \sum_{k=1}^{K} \Vert q^{(k)} \Vert_2^2 - Km\sigma^2 \ge t \right] \hspace{-3pt} + \hspace{-1pt} \PP\left[ \sum_{k=1}^{K} \Vert q^{(k)} \Vert_2 - K\sqrt{m}\sigma \ge t \right] \\
  &\le \exp\left(-c_1\min\left(\frac{t^2}{K m \sigma^4}, \frac{t}{\sigma^2}\right)\right) + \exp(-\frac{c_2 t^2}{K \sigma^4}).
 \end{align*}
 For some constant $c$, setting $t = c\sqrt{K m} \sigma^2 \log^{1/2}(n)$ in the previous inequality implies that the following inequality holds with probability at least $1 - 1/n - 1/n^m$ for any $z \in \mathcal{Z}$.
 \begin{equation}\label{eq:final-tail}
  \frac{1}{K}\sum_{k = 1}^{K} \vert (z^{(k)} - z)^T q^{(k)} \vert \le 4c\sqrt{m}\sigma^2\frac{\log^{1/2}(n)}{\sqrt{K}} + 2m\sigma^2 + 2\sqrt{m}\sigma.
 \end{equation}
 Since $m\sigma^2 < 1$, \refequ{eq:final-tail} implies \refequ{eq:zkqk}.
\end{proof}

\subsection{Proof of \reflemm{lemm:approx-core}}
\label{subsec:proof-lemm-approx-core}
\begin{proof}
 Let $\{w^{(k)}\}$ and $\{q^{(k)}\}$ be sequences generated by \refalg{alg:PPGA}.
 Let $w^* = (x^*, z^*, \gamma^*)$ be a solution to \refequ{eq:modified-opt}.
 Define $\gamma^*_{\max}$ and $\grave{z}$ as:
 \[
  \gamma^*_{\max} = \underset{i, j}{\max} \; \vert \gamma^*_{i,j} \vert \;\; \text{ and } \;\; \grave{z} = \underset{z \in \mathcal{Z}}{\argmax} \; \sum_i \frac{U_i(z) + \upsilon}{U_i(\bar{x}_i) + \upsilon}.
 \]
 Further, define $\varepsilon_1$ and $\varepsilon_2$ as:
 \[
  \varepsilon_1 = \frac{\rho}{K}\sum_{k=1}^{K}\vert (z^{(k)} - \grave{z})^T q^{(k)} \vert + \frac{\rho}{2K},
 \]
 and
 \[
  \varepsilon_2 = \frac{2L\sqrt{n}}{K}\left(\sum_{k=1}^{K}\vert (z^{(k)} - z^*)^T q^{(k)} \vert \right)^{1/2} + \frac{L\sqrt{2n}}{K} + \frac{2L\sqrt{nm}}{\rho K} \gamma^*_{\max} + \upsilon,
 \]
 where $L = (1+\beta)/\sqrt{2}$.
 By \reflemm{lemm:core-1} and the definition of $\grave{z}$, for any $z \in \mathcal{Z}$, we have:
 \begin{equation} \label{eq:core-2}
  \frac{1}{n}\sum_i\frac{U_i(z) + \upsilon}{U_i(\bar{x}_i) + \upsilon} \le \frac{1}{n}\sum_i\frac{U_i(\grave{z}) + \upsilon}{U_i(\bar{x}_i) + \upsilon} \le 1 + \varepsilon_1.
 \end{equation}
 Each $U_i$ is concave, $\beta$-smooth, and bounded between 0 and 1.
 The domain $\mathcal{Z}$ is also bounded (see \refequ{eq:bounded-domain}).
 Therefore, each $U_i$ is $L$-Lipschitz (\refclm{clm:smooth2lips}).
 Given this, and the fact that $\Vert \bar{x}_i - \hat{z} \Vert_2 \le \Vert \bar{x} - G\hat{z} \Vert_2$ for any $i$, \reflemm{lemm:residual} implies the following:
 \begin{equation}
 \label{eq:Lipschitz-1}
  U_i(\bar{x}) \le U_i(\hat{z}) + \varepsilon_2 - \upsilon \quad \quad \forall i.
 \end{equation}
 Combining \refequ{eq:core-2} and \refequ{eq:Lipschitz-1}, for any $z \in \mathcal{Z}$, we have:
 \begin{equation}\label{eq:asymp-core}
  \frac{1}{n}\sum_i \frac{U_i(z)}{U_i(\hat{z}) + \varepsilon_2} \le \frac{1}{n}\sum_i \frac{U_i(z) + \upsilon}{U_i(\hat{z}) + \varepsilon_2} \le 1 + \varepsilon_1.
 \end{equation}

 \noindent
 Given \refequ{eq:asymp-core}, \reflemm{lemm:mnw-core-approx} implies that $\hat{z}$ is an $(\varepsilon_1, \varepsilon_2(1 + \varepsilon_1))$-core outcome.
 Assuming $K = \Theta(n)$, and setting $\alpha = 2\log(1/\delta)/\epsilon + 1$, we have:
 \[
  m\sigma^2 = \frac{K m \alpha}{n^2(\epsilon - \log(1/\delta)/(\alpha - 1))} = \mathcal{O}\left( \frac{m \log(1/\delta)}{n\epsilon^2} \right).
 \]
 Since $m = o(\sqrt{n})$, if we set $\upsilon = \Theta(1/\sqrt{n})$ and choose $\epsilon, \delta > 0$ such that $m\sigma^2 < 1$, then we can apply \reflemm{lemm:zkqk} to establish the lemma.
\end{proof}

\subsection{Proof of \reflemm{lemm:x-step-approx}}
\label{subsec:proof-lemm-x-step-approx}
\begin{proof}
 For any $z, \gamma_i \in \RR^m$, let $h(x_i) \triangleq - L_i^{\rho}(x_i, z, \gamma_i)$.
 To prove the lemma, we rely on \cite[Theorem 4]{lu2023accelerated}.
 To apply this theorem, the following four conditions must hold: (1) $h$ is convex; (2) the diameter of $\mathcal{Z}$ is bounded; (3) there exists a point in the interior of $\mathcal{Z}$; and (4) $h(x_i)$ has a Lipschitz continuous gradient over $\mathcal{Z}$.
 Condition (1) is straightforward to verify.
 Condition (2) is established in \refequ{eq:bounded-domain}.
 For (3), we note that a point such as $z = s / (c \Vert s \Vert)$ lies in the relative interior of $\mathcal{Z}$.
 Regarding (4), by \refclm{clm:log-smooth}, $\theta_i$ has an $L$-Lipschitz continuous gradient over $\mathcal{Z}$ with $L \le \frac{M^2}{\upsilon^2} + \frac{\beta}{\upsilon}$, where $M = \frac{1+\beta}{\sqrt{2}}$.
 It then follows from the definition of $h$ that it has an $(L+\rho)$-Lipschitz continuous gradient over $\mathcal{Z}$.

 \noindent
 Since these four conditions are satisfied, by~\cite[Theorem 4]{lu2023accelerated}, we can find a point $x_i^* \in \mathcal{Z}$ and a residual $v \in \RR^m$ such that:
 \begin{equation}\label{eq:subdiff-ineq}
  v \in \partial \left(h(x_i^*) + \mathbbm{1}_{\mathcal{Z}}(x_i^*) \right), \quad \text{and} \quad \Vert v \Vert_2 \le \frac{\xi}{2},
 \end{equation}
 where $\mathbbm{1}_{S}$ denotes the indicator function%
 \footnote{$\mathbbm{1}_{S}(x)$ equals $0$ if $x \in S$, and $+\infty$ otherwise.}
 of the set $S$, and $\partial f$ denotes the subdifferential%
 \footnote{$\partial f(x) \triangleq \{ u \in \RR^m : f(x\pr) \ge f(x) + u^T(x\pr - x), \forall x\pr \in \text{dom}(f) \}$}
 of the convex function $f$.
 This guarantee can be achieved using at most
 \begin{equation}\label{eq:iteration-bound}
  \mathcal{O} \left( \sqrt{\frac{L+\rho}{\xi}} \log(1/\xi) \right)
 \end{equation}
 projections onto $\mathcal{Z}$, a convergence rate that is optimal up to a logarithmic factor.

 \noindent
 Next, we observe that since $\mathcal{Z}$ has nonempty interior, the subdifferential of the sum of convex functions equals the sum of their subdifferentials~\cite[Theorem 3.40]{beck2017first}.
 Therefore, we can rewrite \refequ{eq:subdiff-ineq} as:
 \begin{equation}\label{eq:subdiff-ineq2}
  v \in \nabla h(x_i^*) + \partial \mathbbm{1}{\mathcal{Z}}(x_i^*), \quad \text{and} \quad \Vert v \Vert_2 \le \frac{\xi}{2}.
 \end{equation}
 It is also straightforward to verify that $\partial \mathbbm{1}_{\mathcal{Z}}(x_i^*) = N_{\mathcal{Z}}(x_i^*)$, where $N_{\mathcal{Z}}$ denotes the normal cone of $\mathcal{Z}$, defined as: $N_S(x) \triangleq \{ y \in \RR^m : y^T(u - x) \le 0, \forall u \in S \}$~\cite{ryu2016primer}.
 Using this identity, we can further rewrite \refequ{eq:subdiff-ineq2} as:
 \begin{equation}\label{eq:subdiff-ineq3}
  v - \nabla h(x_i^*) \in N_{\mathcal{Z}}(x_i^*), \quad \text{and} \quad \Vert v \Vert_2 \le \frac{\xi}{2}.
 \end{equation}
 Finally, since $\sup_{x,x' \in \mathcal{Z}} \Vert x - x' \Vert_2 \le \sqrt{2} \le 2$, we can apply the Cauchy–Schwarz inequality to conclude that, for any $z \in \mathcal{Z}$, the following holds.
 \begin{equation}\label{eq:subdiff-ineq4}
  -(z - x_i^*)^T \nabla h(x_i^*) \le -(z - x_i^{(k)})^T v \le \Vert v \Vert_2 \Vert z - x_i^{(k)} \Vert_2 \le \frac{\xi}{2}(2) = \xi.
 \end{equation}
 Given that $- \nabla h(x_i^*) = \nabla \theta_i(x^*) - \gamma_i - \rho (x_i^* - z)$, it follows that \refequ{eq:subdiff-ineq4} implies \refequ{eq:x-step-approx}.

 \noindent
 It is well-known that projecting a vector in $\RR^m$ onto $\mathcal{Z}$ can be done in $\mathcal{O}(m\log(m))$ time using a sorting-based algorithm (e.g., see \cite[Algorithm 5.1]{hough2024primal}).
 Consequently, a point $x_i^*$ satisfying \refequ{eq:x-step-approx} can be computed in total time
 \[
  \mathcal{O}\left( m \log(m) \cdot \sqrt{ \frac{L+\rho}{\xi}}\log(1/\xi) \right).
 \]
\end{proof}

\end{document}